\documentclass[12pt]{article}

\usepackage{scicite,epsfig,amssymb,amsmath,lscape}

\usepackage{times}
\usepackage{caption}
\usepackage{longtable}

\newcommand{\mnras}{Mon.~Not.~R.~Astron.~Soc.}
\newcommand{\apj}{Astrophys.~J.}
\newcommand{\apjl}{Astrophys.~J.~Lett.}
\newcommand{\swift}{{\it Swift}}
\newcommand{\nat}{{\it Nature}}
\newcommand{\pasp}{{\it Pub. Astron. Soc. Pac.}}
\newcommand{\physrep}{{\it Phys. Rep.}}
\newcommand{\aap}{{\it Astron. Astrophys.}}
\newcommand{\aj}{{\it Astronomical Journal}}
\newcommand{\prd}{{\it Phys. Rev. D}}

\def\arcmin{\hbox{$^\prime$}}
\def\arcsec{\hbox{$^{\prime\prime}$}}

\newdimen\sa  \newdimen\sb
\def\parcs{\sa=.07em \sb=.03em
     \ifmmode $\rlap{.}$^{\scriptscriptstyle\prime\kern -\sb\prime}$\kern -\sa$
     \else \rlap{.}$^{\scriptscriptstyle\prime\kern -\sb\prime}$\kern -\sa\fi}

\newenvironment{sciabstract}{%
\begin{quote} \bf}
{\end{quote}}

\newcounter{lastnote}

\title{Light Curves of the Neutron Star Merger GW170817/SSS17a:\\ Implications for R-Process Nucleosynthesis}

\author
{M.~R.~Drout,$^{1\ast}$
A.~L.~Piro,$^{1}$
B.~J.~Shappee,$^{1,2}$
C.~D.~Kilpatrick,$^{3}$ \\
J.~D.~Simon,$^{1}$
C.~Contreras,$^{4}$
D.~A.~Coulter,$^{3}$
R.~J.~Foley,$^{3}$
M.~R.~Siebert,$^{3}$\\
N.~Morrell,$^{4}$
K.~Boutsia,$^{4}$
F.~Di Mille,$^{4}$
T.~W.-S.~Holoien,$^{1}$
D.~Kasen,$^{5,6}$ \\
J.~A.~Kollmeier,$^{1}$ 
B.~F.~Madore,$^{1}$
A.~J.~Monson,$^{1,7}$ 
A.~Murguia-Berthier,$^{3}$ \\
Y.-C.~Pan,$^{3}$ 
J.~X.~Prochaska,$^{3}$ 
E.~Ramirez-Ruiz,$^{3,8}$ 
A.~Rest,$^{9,10}$ 
C.~Adams,$^{11}$ \\
K.~Alatalo,$^{1,9}$  
E.~Ba\~{n}ados,$^{1}$
J.~Baughman,$^{12,13}$
T.~C.~Beers,$^{14,15}$
R.~A.~Bernstein,$^{1}$ \\
T.~Bitsakis,$^{16}$
A.~Campillay,$^{17}$
T.~T.~Hansen,$^{1}$
C.~R.~Higgs,$^{18,19}$
A.~P.~Ji,$^{1}$ \\
G.~Maravelias,$^{20}$ 
J.~L.~Marshall,$^{21}$ 
C.~Moni Bidin,$^{22}$ 
J.~L.~Prieto,$^{13,23}$ \\
K.~C.~Rasmussen,$^{14,15}$ 
C.~Rojas-Bravo,$^{3}$
A.~L.~Strom,$^{1}$ 
N.~Ulloa,$^{17}$ \\
J.~Vargas-Gonz\'{a}lez,$^{4}$
Z.~Wan,$^{24}$
D.~D.~Whitten$^{14,15}$
\\}

\date{}

\begin{document} 

%Double-space the manuscript.

\baselineskip24pt

% Make the title.

\maketitle 

\noindent
\normalsize{$^{1}$ The Observatories of the Carnegie Institution for Science, 813 Santa Barbara St., Pasadena, CA 91101, USA}\\ 
\normalsize{$^{2}$ Institute for Astronomy, University of Hawai'i, 2680 Woodlawn Drive, Honolulu, HI 96822, USA}\\
\normalsize{$^{3}$ Department of Astronomy and Astrophysics, University of California, Santa Cruz, CA 95064, USA}\\
\normalsize{$^{4}$ Las Campanas Observatory, Carnegie Observatories, Casilla 601, La Serena, Chile}\\
\normalsize{$^{5}$ Departments of Physics and Astronomy, 366 LeConte Hall, University of California, Berkeley, CA, 94720, USA}\\
\normalsize{$^{6}$ Nuclear Science Division, Lawrence Berkeley National Laboratory, Berkeley, CA 94720, USA}\\
\normalsize{$^{7}$ Department of Astronomy \& Astrophysics, The Pennsylvania State University, 525 Davey Lab, University Park, PA 16802, USA}\\
\normalsize{$^{8}$Dark Cosmology Center, Niels Bohr Institute, University of Copenhagen, Blegdamsvej 17, 2100 Copenhagen, Denmark}\\
\normalsize{$^{9}$ Space Telescope Science Institute, 3700 San Martin Drive, Baltimore, MD 21218, USA}\\
\normalsize{$^{10}$ Department of Physics and Astronomy, The Johns Hopkins University, 3400 North Charles Street, Baltimore, MD 21218, USA}\\
\normalsize{$^{11}$ Division of Physics, Mathematics, and Astronomy, California Institute of Technology, Pasadena, CA 91125, USA}\\
\normalsize{$^{12}$ Massachusetts Institute of Technology, Cambridge, MA, USA}\\
\normalsize{$^{13}$ N\'ucleo de Astronom\'ia de la Facultad de Ingenier\'ia y Ciencias, Universidad Diego Portales, Av. Ej \'ercito 441, Santiago, Chile}\\
\normalsize{$^{14}$Department of Physics, University of Notre Dame, Notre Dame, IN 46556, USA}\\
\normalsize{$^{15}$Joint Institute for Nuclear Astrophysics - Center for the Evolution of the Elements, USA}\\
\normalsize{$^{16}$ Instituto de Radioastronom\'{i}a y Astrof\'{i}sica, Universidad Nacional Aut\'{o}noma de M\'{e}xico, C.P. 58190, Morelia, Mexico}\\
\normalsize{$^{17}$Departamento de F\'{i}sica y Astronom\'{i}a, Facultad de Ciencias, Universidad de La Serena, Cisternas 1200, La Serena, Chile}\\
\normalsize{$^{18}$University of Victoria, Victoria, B.C., Canada}\\
\normalsize{$^{19}$National Research Council Herzberg Institute of Astrophysics, 5071 West Saanich Road, Victoria, B.C., V9E 2E7, Canada}\\
\normalsize{$^{20}$ Instituto de F\'{i}sica y Astronom\'{i}a, Universidad de Valpara\'{i}so, Av. Gran Breta\~na 1111, Casilla 5030, Valpara\'iso, Chile}\\
\normalsize{$^{21}$George P. and Cynthia Woods Mitchell Institute for Fundamental Physics and Astronomy, and Department of Physics and Astronomy, Texas A\&M University, College Station, TX 77843, USA}\\
\normalsize{$^{22}$Instituto de Astronom\'{i}a, Universidad Cat\'{o}lica del Norte, Av. Angamos 0610, Antofagasta, Chile}\\
\normalsize{$^{23}$Millennium Institute of Astrophysics, Santiago, Chile}\\
\normalsize{$^{24}$ Sydney Institute for Astronomy, School of Physics, A28, University of Sydney, NSW 2006, Australia}\\

\normalsize{$^\ast$To whom correspondence should be addressed; E-mail:  mdrout@carnegiescience.edu.}

\baselineskip24pt
\begin{sciabstract}

On 2017 August 17, gravitational waves were detected from a binary neutron star merger, GW170817, along with a coincident short gamma-ray burst, GRB170817A. An optical transient source, Swope Supernova Survey 17a  (SSS17a), was subsequently identified as the counterpart of this event. We present ultraviolet, optical and infrared light curves of SSS17a extending from 10.9 hours to 18 days post-merger. We constrain the radioactively-powered transient resulting from the ejection of neutron-rich material. The fast rise of the light curves, subsequent decay, and rapid color evolution are consistent with multiple ejecta components of differing lanthanide abundance. The late-time light curve indicates that SSS17a produced at least $\sim$0.05 solar masses of heavy elements, demonstrating that neutron star mergers play a role in r-process nucleosynthesis in the Universe.

\end{sciabstract}

The discovery of gravitational waves (GWs) from coalescing binary black holes by the Laser Interferometer Gravitational Wave Observatory (LIGO) has transformed the study of compact objects in the Universe \cite{Abbott16:bbh,Abbott16:gw15}. Unlike black holes, merging neutron stars are expected to produce electromagnetic radiation. The electromagnetic signature of such an event can provide more information than the GW signal alone: constraining location of the source, reducing the degeneracies in GW parameter estimation \cite{Hughes2003}, probing the expansion rate of the Universe \cite{Holz2005,Nissanke2013}, and producing a more complete picture of the merger process \cite{Phinney2009,Mandel2010}.

Short gamma-ray bursts (GRBs) have long been expected to result from neutron star mergers \cite{Paczynski1986,Eichler1989}, and therefore would be a natural electromagnetic counterpart to GWs\cite{Kelley2013}. Unfortunately, their emission is beamed, so that it may not intersect our line of sight\cite{Fong2015}. The possibility that only a small fraction of GRBs may be detectable has motivated theoretical and observational searches for more-isotropic electromagnetic signatures, such as an astronomical transient powered by the radioactive decay of neutron-rich ejecta from the merger. \cite{Li1998,Kulkarni2005,Metzger2010,Roberts2011,Piran2013,Metzger2017}. Referred to as a macronova or kilonova, the detection of these events would provide information on the origin of many of the heaviest elements in the periodic table\cite{Shen2015}. 

It has long been realized that approximately half of the elements heavier than iron are created via r-process nucleosynthesis---the capture of neutrons onto lighter seed nuclei on a timescale more rapid than $\beta$-decay pathways \cite{Burbidge1957,Cameron1957}. However, it is less clear where the r-process predominantly occurs, namely whether the primary sources of these elements are core-collapse supernovae or compact binary mergers (black hole--neutron star or neutron star--neutron star) \cite{Qian2007,Arnould2007}. For supernovae, direct detection of the electromagnetic signatures from r-process nucleosynthesis is obscured by the much larger luminosity originating from hydrogen recombination (for hydrogen-rich supernovae) or nickel-56 and cobalt-56 decay (for hydrogen-poor supernovae). By contrast, it may be possible to measure the r-process nucleosynthesis after a compact object merger from the associated transient, based on its radioactive decay. Such a measurement would demonstrate directly that r-process elements are produced in compact mergers, and provide an estimate of the r-process yield. While there has been some tentative evidence for kilonovae following short gamma-ray bursts \cite{Tanvir2013,Berger2013}, no conclusive event has yet been observed. 

On 2017 August 17, LIGO and Virgo detected the gravitational wave source GW170817, which was identified as a binary neutron star merger based on the waveform \cite{GCN21509,GCN21513,Abbott17:ns}. At 23:33 UTC on 2017 August 17 (10.86 hours post-merger), an optical transient, Swope Supernova Survey 2017a (SSS17a), was identified in the galaxy NGC 4993 by the 1M2H collaboration and was determined to be associated with this event \cite{GCN21529,Coulter2017}. Within an hour of the identification, we began observing the spectral energy distribution (SED) of SSS17a from the $g$- to $K$-bands with the Magellan telescopes\cite{GCN21551}. Early spectra of the source, also obtained within an hour of the optical discovery, were blue and smooth, indicating that the transient event was initially very hot \cite{GCN21547,Shappee2017}. Over the following weeks, we acquired optical and near-infrared (near-IR) imaging of SSS17a at Las Campanas Observatory and W. M. Keck Observatory with the Swope, du Pont, Magellan, and Keck-I telescopes, which are analyzed below \cite{MM}. A companion paper presents optical spectroscopy of SSS17a for an overlapping time period \cite{Shappee2017}. Figure 1A shows the discovery image, composed of $g$-, $i$-, and $H$-band Magellan/Swope imaging from the night of August 17. For comparison, Figure 1B shows a color image from observations obtained 4 days later. The change in color of SSS17a between these images demonstrates the rapid evolution of this transient.

The resulting light curves are shown in Figure 2, augmented with measurements made from public {\it Swift} imaging at ultraviolet wavelengths, and European Southern Observatory (ESO) images in the optical and near-IR \cite{MM}. SSS17a undergoes a rapid rise on a timescale that varies with wavelength, from $<$12 hours in the ultraviolet (UV) and optical bands, to 1--2 days in the near-IR. Over subsequent days, the transient fades quickly. This decline proceeds most rapidly in the bluest bands, where SSS17a fades by $\gtrsim$ 1.5 mag day$^{-1}$, but more slowly in the near-IR, where a $\sim$ 3.5 magnitude decline takes nearly 3 weeks. After correcting for foreground Milky Way reddening \cite{MM} and the distance to NGC 4993 of 39.5 megaparsecs (Mpc), we find that SSS17a has a peak magnitude of $-16.04$ mag in the optical ($V-$band) and $-15.51$ mag in the near-IR ($H-$band), and undergoes a large color evolution. Between 0.5 and 4.5 days post-merger the $V-H$ color of SSS17a transitions from $-$1.2 mag to 3.6 mag (Figure S1). While SSS17a reaches absolute magnitudes typically associated with faint core-collapse supernovae, it both declines in magnitude and evolves to redder colors more rapidly than known optical extragalactic transients \cite{Siebert17,MM}.

We construct UV to near-IR SEDs for SSS17a at ten epochs between 0.5 and 8.5 days after the gravitational wave trigger (Figure 3). Within eight days, the peak of the SED falls by a factor of $\gtrsim$70 in flux, and shifts from the near-UV ($\lesssim$4500~\AA{}) to the near-IR ($\gtrsim$1.5~$\mu$m). The SED at each epoch can be fitted with a blackbody distribution (reduced $\chi^2\sim1$--2), so we consider that the emission is largely thermal. Some deviations are present, most notably an excess around 1~$\mu$m ($Y$-band) present from day 1.5 onward \cite{MM}. The associated color temperatures show that between twelve and thirty-six hours post-merger, SSS17a cooled from $\sim$10,000$\,{\rm K}$ to $\sim$5,100$\,{\rm K}$. Between 0.5 and 5.5 days post-merger, the evolution of the color temperature ($T_{\rm c}$) with time ($t$) is consistent with a power-law decline: $T_{\rm c}$ $\propto$ $t^{-0.54\pm0.01}$. After 5.5 days, the temperature asymptotically approached $\sim$2500$\,{\rm K}$.

Using the SEDs from each epoch shown in Figure 3, we construct a pseudo-bolometric light curve, which accounts for flux across the electromagnetic spectrum. We compute and sum the SED fluxes using an iterative technique \cite{MM}. To account for flux outside the range of our observations, we extrapolate blackbody emission based on our best-fitting distributions. For flux at shorter wavelengths than our data, the correction factor is $\sim$40\% at 0.5 days---as the temperature is hottest and our observations are limited to wavelengths $\lambda \gtrsim$ 4500 \AA\---but it falls below 1\% by 0.67 days, when \emph{Swift-}UVOT observations begin and the transient rapidly cools. The complementary correction factor for flux at longer wavelengths than our $K$-band observations ranges from $\sim$1\% at day 0.5 to 38\% at day 8.5. We plot the resulting pseudo-bolometric light curve in Figure 4A. The lower limits of the error bars show the amount of flux that we directly observed. In Figure 4C, we combine our fitted temperatures with the bolometric luminosity ($L_{\rm{bol}}$) to estimate an effective photospheric radius ($R_{\rm{phot}}$). 

For observations $>$8.5 days after the merger, we only detect the source in the either the {\it H}- or {\it K}-band at any given time, so we cannot directly measure the temperature. To estimate bolometric luminosities and photospheric radii at these later epochs, we assume an effective temperature of $2,500^{+500}_{-1000}\,{\rm K}$. Though the physical motivation for this choice is further detailed below, observationally, the measurable color temperature is approaching this value from 5.5--8.5 days post-merger. Further, the {\it H}- and {\it K}-bands fall near the peak of the SED for blackbodies in the temperature range 1500--3000~K. As a result, bolometric corrections for either the {\it H}- or {\it K}-band over this entire temperature range lead to a variation in the estimated luminosity of less than a factor of 1.6. Error bars representing this full range are included in Figure 4A and 4C.

The pseudo-bolometric light curve has a peak value of $\sim$10$^{42}$ erg s$^{-1}$ at 0.5 days post-merger, corresponding to our first epoch of observations, and the total radiated energy over 18 days is $\sim$1.7 $\times$ 10$^{47}$ erg. Between 0.5 and 5.5 days post-merger, the bolometric light curve is consistent with a power-law decline of $L_{\rm bol}$ $\propto$ $t^{-0.85 \pm 0.01}$. After 5.5 days, the best-fitting power-law is steeper, with $L_{\rm bol}$ $\propto$ $t^{-1.33 \pm 0.15}$ between 7.5 and 13.5 days.

We use the evolution of $L_{\rm bol}$, $T_{\rm c}$, and $R_{\rm phot}$ to constrain the energy source powering the emission from SSS17a. We first explore whether the physical properties of SSS17a are consistent with a transient powered by the radioactive decay of r-process elements. Models for r-process powered transients predict that the energy generation rate, $\dot{q}_r$, is proportional to $t^{-1.3}$ \cite{Metzger2010,Roberts2011,MM}. This power-law is similar to the slope observed in the late-time bolometric light curve of SSS17a. To directly compare the predictions for r-process heating to our observed luminosities, we multiply this intrinsic heating rate by a time-dependent thermalization efficiency \hbox{(60--25\%)}\cite{MM}, and fit our data. According to Arnett's Law, the peak luminosity of a radioactively-powered transient should correspond to the instantaneous heating rate \cite{Arnett1982}. Under the hypothesis that the luminosity at 0.5 days post-merger is due to r-process heating, this implies that $\sim$0.01 solar masses ($\,M_\odot$) of r-process material was generated. The heating rate for this mass of r-process material, M$_{\rm{r-p}}$, is plotted in Figure 4A.

While heating from $\sim$0.01$\,M_\odot$ of r-process ejecta could explain the peak observed luminosity, it would have several further consequences. First, the fast rise ($<$0.5 days) would require that the specific opacity, $\kappa$, of this material be less than $\sim$0.08$\,{\rm cm^2\,g^{-1}}$ \cite{MM}. The opacity is strongly dependent on the presence of lanthanide elements, because they have a large number of bound-bound transitions due to the presence of an open f shell \cite{Kasen2013}. This low inferred opacity would thus imply that the early ejecta cannot be lanthanide-rich. Then, the abundance of lanthanides is strongly dependent on the neutron-richness of the ejecta, often expressed as the electron fraction $Y_{e}$, where $Y_{e} = 0.5$ for symmetric matter (equal proportions of neutrons and protons) and $Y_{e} = 0$ for pure neutrons. To produce material with such low opacity that is relatively lanthanide-free would require $Y_{e} \gtrsim 0.3$. 

Second, this low inferred opacity would cause the associated material to quickly become optically thin (within $\sim$2 days; when SSS17a is blue/hot). A low optical depth is inconsistent with the continuing optical emission that we observed over the following weeks from SSS17a, so this model necessitates an additional higher-opacity component. Comparing the r-process heating to the later light curve yields a mass estimate of $0.05 \pm 0.02\,M_\odot$ (Figure 4A), but for SSS17a to remain optically thick for a timescale of 2--3 weeks requires an opacity \hbox{$\kappa\gtrsim5\,{\rm cm^3\,g^{-1}}$.} The evolution of the light curve over this time interval therefore constitutes evidence for a second, lanthanide-rich component, which dominates at later times when the SSS17a is red/cool.

Such two-component ejecta are generally expected for neutron star mergers \cite{Perego2014,Fernandez2016}. This structure could correspond to two distinct physical components, where the lanthanide-rich component arises from material ejected on dynamical timescales via processes such as tidal forces \cite{Hotokezaka2013} and the lanthanide-free component forms on longer timescales ($\sim$seconds), such as from  the accretion disk wind \cite{Fernandez2015}. Alternatively, both of these compositional components could arise from the same dynamical ejecta \cite{Wanajo2014,Bovard17}. The exact contribution of each component to the observed light curve depends on the mass ratio of the merging binary, as well as the orientation relative to the line of sight \cite{Kasen2015}. For example, it is possible that the blue component could be underestimated if it is partially obscured/absorbed by the material producing the red component. Detailed modeling, which accounts for these degeneracies, is presented in a companion paper \cite{Kilpatrick17}.

Figure 4C shows the evolution of the measured radii. A comparison to model curves for material moving at 10\%, 20\%, and 30\% of the speed of light indicates that the photosphere expands at relativistic speeds in the first few days. However, after about 5 days, the photosphere begins moving inward. This behavior is reminiscent of hydrogen-rich core-collapse supernovae following hydrogen recombination \cite{Elmhamdi2003}, and a similar process may be occurring here. In the case of an r-process powered transient, recombination of the open f-shell lanthanide elements, such as neodymium, is expected to begin at a temperature of $\sim$2,500$\,{\rm K}$ \cite{Kasen2013}. These ionized elements are the dominant opacity source, so the recombination causes the opacity to decline rapidly and the photosphere to move inward. This interpretation is corroborated by the effective temperature of $\sim$2,500$\,{\rm K}$ that we measure from the SED for $t > 5$\,days, and supports our assumption of a roughly constant temperature throughout the remainder of the evolution. 

Other processes have been considered for providing an optical counterpart to neutron star mergers, including magnetic dipole spin down, heating from radioactive nickel, and cocoon emission e.g. \cite{Metzger14,Metzger2009,Gottlieb17}. These models must be compared with our detailed observations as well. For instance, luminosity powered by the spin-down of a magnetic dipole is predicted to scale as $L_{\rm{bol}} \propto t^{-2}$, steeper than the measured bolometric light curve of SSS17a, and should produce strong X-ray emission \cite{Metzger14}. Then, similar to r-process heating, power from radioactive nickel cannot self-consistently reproduce the entire photometric evolution of SSS17a---fitting both the peak luminosity and fast decline leads to the unphysical requirement that the mass of radioactive nickel approaches or exceeds the total ejecta mass \cite{MM}. Still, we find that it is possible to reproduce the bolometric evolution between 7.5 and 18 days post-merger with heating due to $\sim$0.002 M$_\odot$ of radioactive nickel \cite{MM}---if another emission process dominates at early times. However, nickel heating due does not naturally explain the temperature evolution observed in SSS17a. A rapid evolution to very red colors is not observed in other known transients powered by radioactive nickel \cite{Siebert17}.

Thus, we conclude that the late-time ($\gtrsim$ 5 days) decay rate and color evolution of SSS17a are consistent with a transient powered by the radioactive decay of r-process elements. If the early emission is also powered by r-process heating, multiple ejecta components with differing lanthanide abundances are required. Overall, we estimate that at least $\sim$0.05$\,M_\odot$ of r-process material is generated in this event from the late-time light curve. 

The predicted mass fraction of lanthanides in this material is $\sim$0.1--0.5, depending on $Y_e$ \cite{Wanajo2014}. Typical solar abundance (by mass fraction) for the r-process elements with mass number $A>100$ is $\sim8\times10^{-8}$ \cite{Kappeler89}, resulting in a Milky Way r-process production rate of $\sim3\times10^{-7}\,M_\odot\,{\rm yr^{-1}}$ \cite{Qian2000,Metzger2009}. If neutron star mergers dominate r-process production, this production rate requires an event like GW170817/SSS17a in our Galaxy every 20,000--80,000 years, or a volume density of $\sim$(1--4)$\times10^{-7}\,{\rm Mpc^{-3}\,yr^{-1}}$. At their design sensitivity, Advanced LIGO, Advanced Virgo and the Kamioka Gravitational Wave Detector (KAGRA) will be able to detect binary neutron star mergers out to 200 Mpc \cite{Abbott16:review}, leading to a possible detection rate of $\sim$3$-$12 per year. This rate translates to less than one event per year as nearby as GW170817/SSS17a. This number would increase if the r-process mass we calculate for SSS17a is overestimated. Such an overestimate could occur if our assumed heating efficiency is too low or if this event produced more ejecta than an average neutron star merger. 

Empirical explanation for the portion of the periodic table expected to result from r-process nucleosynthesis has been elusive. The UV to near-IR light curves of the neutron star merger GW170817/SSS17a provide evidence for binary neutron star mergers as an origin for these elements. Observations of more events are now required to precisely map r-process yields from this channel.

\clearpage
\begin{figure}
\begin{center}
\includegraphics*[width=0.95\textwidth]{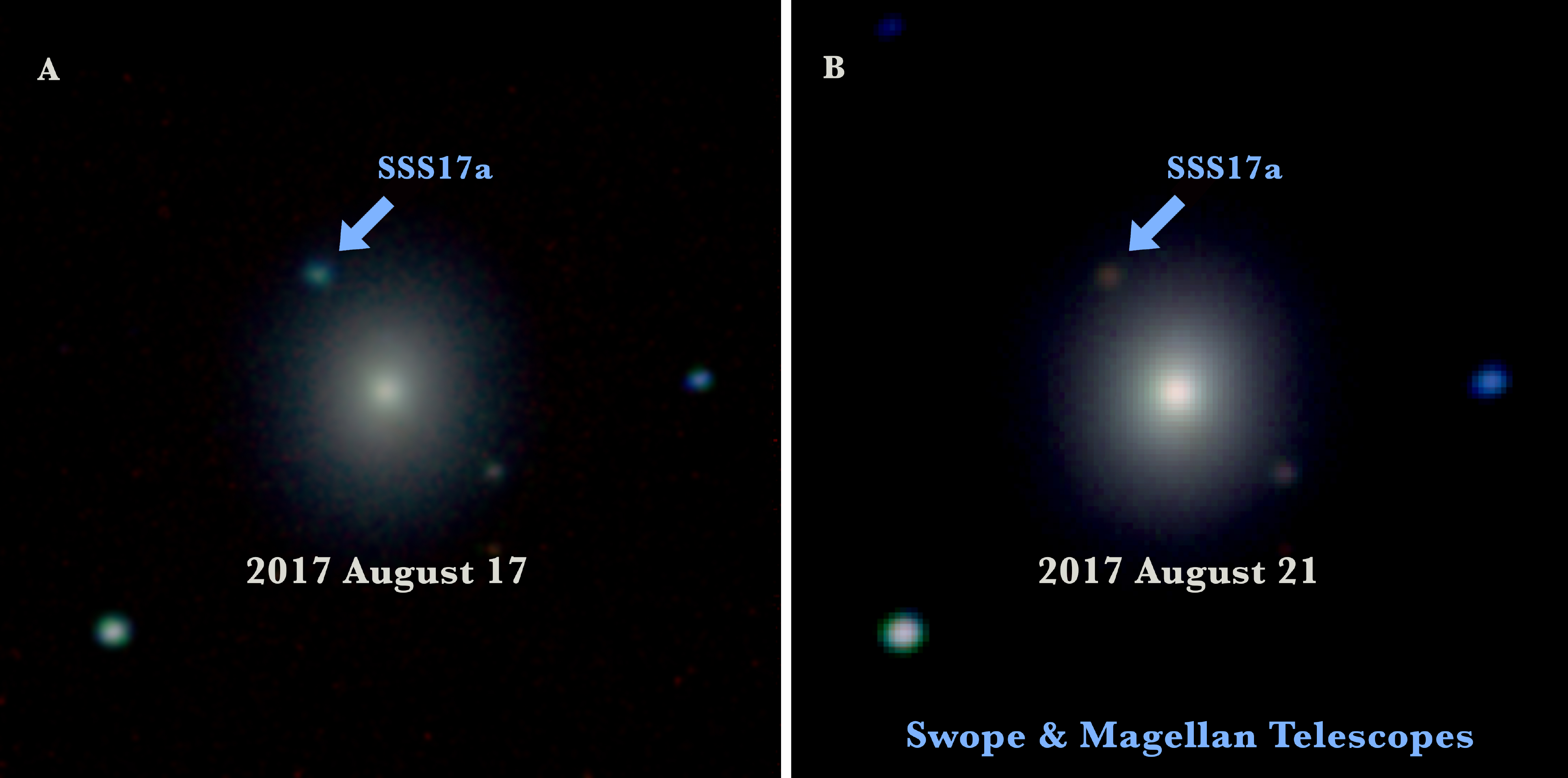}
\caption{{\bf Pseudo-color images of SSS17a in the galaxy NGC\,4993.} 
Images are 1\arcmin\ $\times$ 1\arcmin\ and centered on NGC\,4993; SSS17a is indicated by a blue arrow in each panel. The red, green, and blue channels correspond to the $H$-band, $i$-band, and $g$-band images described in \cite{MM}. {\bf (A)} Images taken on the night of 2017 August 17, 0.5 days after the merger. {\bf (B)} Images taken on the night of 2017 August 21, 4.5 days after the merger. Over four days SSS17a both faded and became redder.}
\end{center}
\end{figure}

\clearpage

\begin{figure}
\begin{center}
\includegraphics*[width=0.8\textwidth]{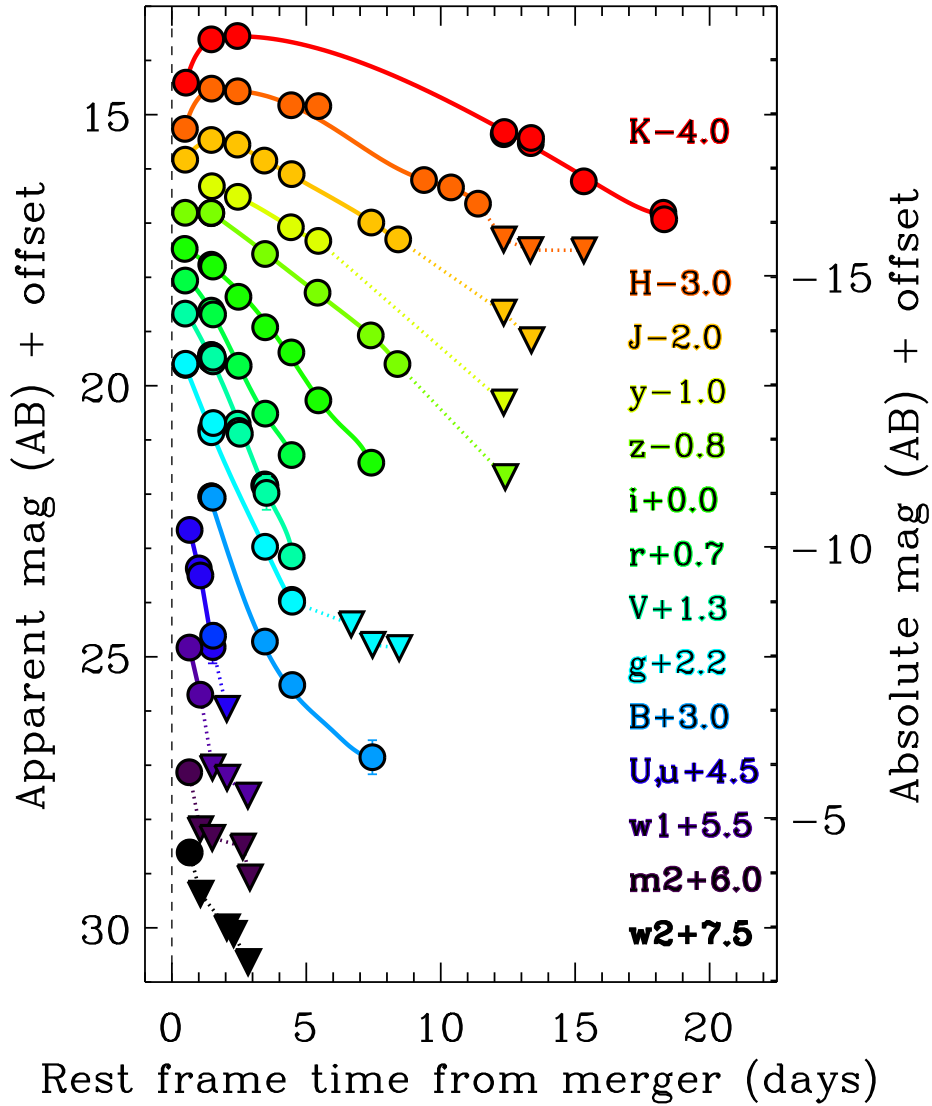}
\caption{{\bf Ultraviolet to near-infrared photometry of SSS17a.} Observations begin 10.9 hours after merger and continue to +18.5 rest-frame days. SSS17a exhibits both a rapid rise and decline, and becomes substantially redder with time. Detections are shown as circles and connected by solid lines for a given photometric band. Upper limits are shown as triangles and connected by dotted lines. The time of merger is indicated by a vertical dashed line. The right hand vertical axis accounts only for the distance to the host galaxy, NGC\,4993. For absolute magnitudes corrected for foreground Milky Way reddening, see \cite{MM}}
\end{center}
\end{figure}

\clearpage

\begin{figure}
\begin{center}
\includegraphics*[width=0.7\textwidth,trim=0 0 0 1cm, clip]{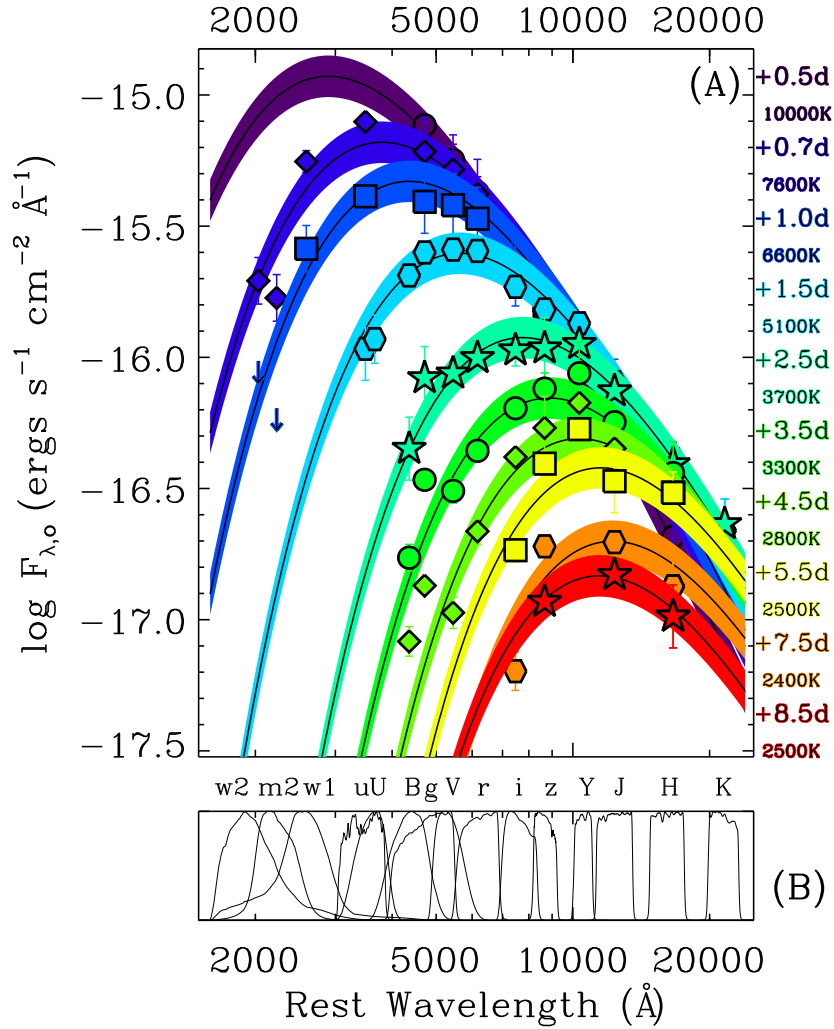}
\caption{{\bf Evolution of the ultraviolet to near-infrared spectral energy distribution (SED) of SSS17a.} {\bf (A)} The vertical axis, $\log$ F$_{\lambda,\rm{o}}$, is the logarithm of the observed flux. Fluxes have been corrected for foreground Milky Way extinction \cite{MM}. Detections are plotted as filled symbols and upper limits for the third epoch (1.0 days post-merger) as downward pointing arrows. Less-constraining upper limits at other epochs are not plotted for clarity. Between 0.5 and 8.5 days after the merger, the peak of the SED shifts from the near-UV ($<$4500~\AA\,) to the near-IR ($>$1$~\mu$m), and fades by a factor $>$70. The SED is broadly consistent with a thermal distribution and the colored curves represent best-fitting blackbody models at each epoch. In 24 hours after the discovery of SSS17a, the observed color temperature falls from $\gtrsim$10,000 K to $\sim$5,000 K. The epoch and best-fitting blackbody temperature (rounded to 100 K) are listed. SEDs for each epoch are also plotted individually in Figure S2 and described in \cite{MM}. {\bf (B)} Filter transmission functions for the observed photometric bands.}
\end{center}
\end{figure}

\clearpage

\begin{figure}
\begin{center}
\includegraphics*[width=0.6\textwidth,trim=0 3cm 0 0, clip]{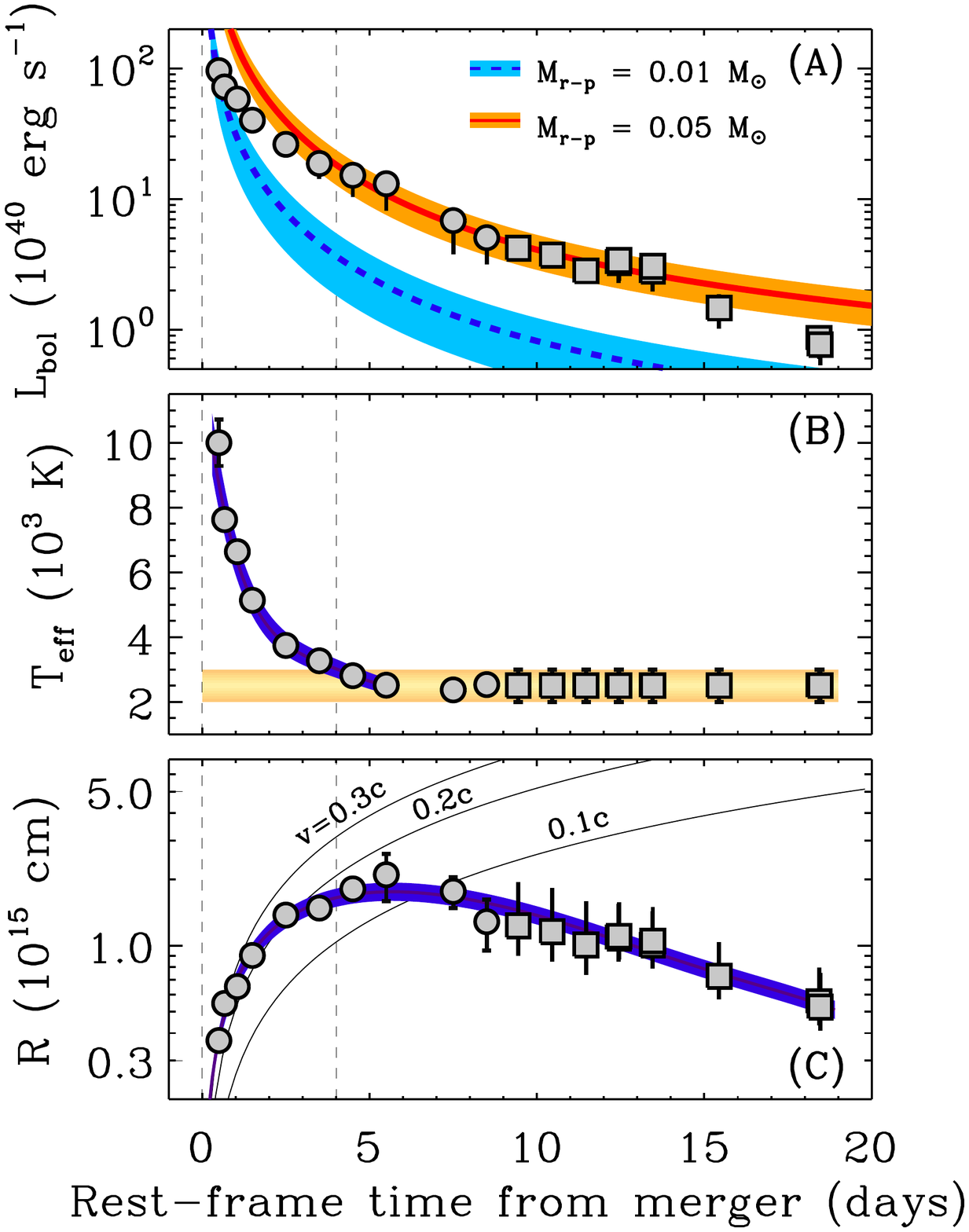}
\caption{{\bf Physical parameters derived from the ultraviolet to near-infrared SEDs of SSS17.} Vertical dashed lines indicate the time of merger and four days post-merger, between which SSS17a undergoes a period of rapid expansion and cooling. {\bf (A)} Pseudo-bolometric light curve evolution; representative r-process radioactive heating curves are also shown. While the initial observed peak is consistent with $\sim$0.01~M$_\odot$ of r-process material (blue curve), this under-predicts the luminosity at later times. Instead, the late-time ($>$ 4 day) light curve matches radioactive heating from $0.05 \pm 0.02$~M$_\odot$ of r-process material (red curves).
{\bf (B)} Best-fitting blackbody model temperatures. 11 hours after the merger, SSS17a is consistent with a blackbody of $\gtrsim$10,000K. Between 4.5 and 8.5 days, the temperature asymptotically approaches $\sim$2500K --- the temperature at which open f-shell lanthanide elements are expected to recombine. Radii and luminosities beyond 8.5 days are computed assuming a temperature of $2500^{+500}_{-1000}$~K and are plotted as squares. This temperature range is highlighted by the orange horizontal band. 
{\bf (C)} Best-fitting blackbody model radii. Curved lines represent the radius of material moving at 10\%, 20\%, and 30\% the speed of light. At early times the increase in radius with time implies that the ejecta are expanding relativistically. After $\sim$5~days, the measured radii decrease, likely due to recombination.}
\end{center}
\end{figure}

\clearpage

%\bibliography{gw}
%\bibliographystyle{Science}

%\begin{scilastnote}
\section*{Acknowledgements}

We thank John Mulchaey (Carnegie Observatories director), Leopoldo Infante (Las Campanas Observatory director), and the entire Las Campanas staff for their dedication, professionalism, and excitement, which were all critical in obtaining the observations used in this study. We also thank Ian Thompson and the Carnegie Observatory Time Allocation Committee for approving the Swope Supernova Survey and scheduling our program. We thank the University of Copenhagen, DARK Cosmology Centre, and the Niels Bohr International Academy for hosting D.A.C., R.J.F., A.M.B., E.R., and M.R.S.\ during this work.  R.J.F., A.M.B., and E.R.\ were participating in the Kavli Summer Program in Astrophysics, ``Astrophysics with gravitational wave detections.''  This program was supported by the the Kavli Foundation, Danish National Research Foundation, the Niels Bohr International Academy, and the DARK Cosmology Centre.

M.R.D., B.J.S., K.A.A., and A.P.J. were supported by NASA through Hubble Fellowships awarded by the Space Telescope Science Institute, which is operated by the Association of Universities for Research in Astronomy, Inc., for NASA, under contract NAS 5-26555. M.R.D. is a Hubble and Carnegie-Dunlap Fellow. M.R.D. acknowledges support from the Dunlap Institue at the University of Toronto, and thanks M.W.B.~Wilson, L.Z.~Kelly, C.~McCully, and R.~Margutti for helpful discussions. 

The UCSC group is supported in part by NSF grant AST--1518052, the Gordon \& Betty Moore Foundation, the Heising-Simons Foundation, generous donations from many individuals through a UCSC Giving Day grant, and from fellowships from the Alfred P.\ Sloan Foundation (R.J.F), the David and Lucile Packard Foundation (R.J.F.\ and E.R.) and the Niels Bohr Professorship from the DNRF (E.R.). D.K. is supported in part by a Department of Energy (DOE) Early Career award DE-SC0008067, a DOE Office of Nuclear Physics award DE-SC0017616, and a DOE SciDAC award DE-SC0018297, and by the Director, Office of Energy Research, Office of High Energy and Nuclear Physics, Divisions of Nuclear Physics, of the U.S. Department of Energy under Contract No.DE-AC02-05CH11231. 

Support for J.L.P. is in part provided by FONDECYT through the grant 1151445 and by the Ministry of Economy, Development, and Tourism's Millennium Science Initiative through grant IC120009, awarded to The Millennium Institute of Astrophysics, MAS. C.M.B. was supported by FONDECYT through regular project 1150060. G.M. acknowledges support from CONICYT, Programa de Astronom\'ia/PCI, FONDO ALMA 2014, Proyecto No 31140024. A.M.B.\ acknowledges support from a UCMEXUS-CONACYT Doctoral Fellowship. CA was supported by Caltech through a Summer Undergraduate Research Fellowship (SURF) with funding from the Associates SURF Endowment. T.C.B., K.C.R., and D.D.W. 
acknowledge partial support for this work from grant PHY 14-30152; Physics Frontier Center/JINA Center for the Evolution of the Elements (JINA-CEE), awarded by the US National Science Foundation, and from the Luksic Foundation.

This paper includes data gathered with the 6.5 meter Magellan Telescopes located at Las Campanas Observatory, Chile.  This work is based in part on observations collected at the European Organisation for Astronomical Research in the Southern Hemisphere, Chile as part of PESSTO (the Public ESO Spectroscopic Survey for Transient Objects Survey) through ESO program 199.D-0143.
Some of the data presented herein were obtained at the W.\ M.\ Keck Observatory, which is operated as a scientific partnership among the California Institute of Technology, the University of California, and the National Aeronautics and Space Administration. The Observatory was made possible by the generous financial support of the W.\ M.\ Keck Foundation.  The authors wish to recognize and acknowledge the very significant cultural role and reverence that the summit of Maunakea has always had within the indigenous Hawaiian community.  We are most fortunate to have the opportunity to conduct observations from this mountain. This research has made use of the NASA/IPAC Extragalactic Database (NED) which is operated by the Jet Propulsion Laboratory, California Institute of Technology, under contract with the National Aeronautics
and Space Administration. This publication makes use of data products from the Two Micron All Sky Survey, which is a joint project of the University of Massachusetts and the Infrared Processing and Analysis Center/California Institute of Technology, funded by the National Aeronautics and Space Administration and the National Science Foundation.

The data presented in this work and code used to perform the analysis is available at ftp://ftp.obs.carnegiescience.edu/pub/SSS17a. ESO and \emph{Swift-}UVOT data analyzed in this work are available at {http://archive.eso.org/eso/eso\_archive\_main.html} (program ID 199.D-0143) and \\ {https://archive.stsci.edu/swiftuvot/search.php} (target IDs 12167, 12978, and 12979), respectively. Reduced photometry is presented in Table S1 and is also avaliable at WISeREP \cite{Yaron2012} (https://wiserep.weizmann.ac.il/) and on the Open Supernova Catalog \cite{Guillochon2017} (sne.space).

\clearpage

\noindent {{\bf Supplemetary Materials:}\\
{\tt www.sciencemag.org}\\
Materials and Methods \\
%\indent \indent \indent Supplementary text\\
Figures~S1, S2, S3\\
Tables~S1, S2\\
References ($54$--$89$)}

%\end{scilastnote}

%\end{document}

\clearpage
\setcounter{page}{1}
\setcounter{figure}{0}    
\renewcommand{\thefigure}{S\arabic{figure}}
\renewcommand{\thetable}{S\arabic{table}}
\renewcommand{\thesection}{S\arabic{section}}
\renewcommand{\theequation}{S\arabic{equation}}

\begin{center}
\title{{\LARGE Supplementary Materials for}\\[0.5cm]
{\bf\large{Light Curves of the Neutron Star Merger GW170817/SSS17a:\\ Implications for R-Process Nucleosynthesis}}}

\author
{M.~R.~Drout,$^{1\ast}$
A.~L.~Piro,$^{1}$
B.~J.~Shappee,$^{1,2}$
C.~D.~Kilpatrick,$^{3}$ \\
J.~D.~Simon,$^{1}$
C.~Contreras,$^{4}$
D.~A.~Coulter,$^{3}$
R.~J.~Foley,$^{3}$
M.~R.~Siebert,$^{3}$\\
N.~Morrell,$^{4}$
K.~Boutsia,$^{4}$
F.~Di Mille,$^{4}$
T.~W.-S.~Holoien,$^{1}$
D.~Kasen,$^{5,6}$ \\
J.~A.~Kollmeier,$^{1}$ 
B.~F.~Madore,$^{1}$
A.~J.~Monson,$^{1,7}$ 
A.~Murguia-Berthier,$^{3}$ \\
Y.-C.~Pan,$^{3}$ 
J.~X.~Prochaska,$^{3}$ 
E.~Ramirez-Ruiz,$^{3,8}$ 
A.~Rest,$^{9,10}$ 
C.~Adams,$^{11}$ \\
K.~Alatalo,$^{1,9}$  
E.~Ba\~{n}ados,$^{1}$
J.~Baughman,$^{12,13}$
T.~C.~Beers,$^{14,15}$
R.~A.~Bernstein,$^{1}$ \\
T.~Bitsakis,$^{16}$
A.~Campillay,$^{17}$
T.~T.~Hansen,$^{1}$
C.~R.~Higgs,$^{18,19}$
A.~P.~Ji,$^{1}$ \\
G.~Maravelias,$^{20}$ 
J.~L.~Marshall,$^{21}$ 
C.~Moni Bidin,$^{22}$ 
J.~L.~Prieto,$^{13,23}$ \\
K.~C.~Rasmussen,$^{14,15}$ 
C.~Rojas-Bravo,$^{3}$
A.~L.~Strom,$^{1}$ 
N.~Ulloa,$^{17}$ \\
J.~Vargas-Gonz\'{a}lez,$^{4}$
Z.~Wan,$^{24}$
D.~D.~Whitten$^{14,15}$
\\}

\normalsize{Correspondence to: mdrout@carnegiescience.edu.}

\end{center}

{{\bf This PDF file includes:}\\
%{\tt www.sciencemag.org}\\
\indent \indent \indent Materials and Methods \\
%\indent \indent \indent Supplementary text\\
\indent \indent \indent Figures~S1, S2, S3\\
\indent \indent \indent Tables~S1, S2\\
\indent \indent \indent References ($54$--$89$)}

\clearpage

\noindent {\bf \LARGE Materials and Methods}

\section{Data Acquisition \& Reductions}

Swope Supernovae Survey 17a (SSS17a) was discovered in an $i$-band image obtained with the 1-m Swope telescope at Las Campanas Observatory at UT 2017 August 17 23:33, 10.9 hours after the LIGO Scientific Collaboration and Virgo Collaboration (referred to jointly as LVC) gravitational wave trigger \cite{GCN21529,Coulter2017}. Immediately after discovery, we initiated a follow-up campaign of optical and near-IR photometric observations of SSS17a spanning from $\sim$11 hours to 18.5 days after the LVC trigger. In the sections below, we describe the data acquisition, reduction, and calibration. All photometry is presented in Table S1 on the AB magnitude scale.

\subsection{Swope Photometry and Field Star Calibration}

Following the discovery of SSS17a, we obtained observations with the Swope telescope in the \emph{BVgri} bands on five additional nights spanning 2017 August 18 to 2017 August 24. Details of the observations and data reduction are described in a companion paper \cite{Coulter2017}
We performed all reductions of our Swope imaging using {\tt photpipe} \cite{Rest2005,Rest2014} as described in \cite{Coulter2017}. Panoramic Survey Telescope and Rapid Response System (PanSTARRS)
magnitudes of stars in the field of SSS17a were transformed into the Swope natural system using Supercal transformations \cite{Scolnic15}.

\subsection{Magellan/LDSS-3 and Magellan/IMACS Optical Imaging}

Photometric observations of SSS17a post-discovery included a series of three, 30~s $g$-band images obtained with the LDSS-3 imaging spectrograph on the 6.5-m Magellan/Clay telescope \cite{GCN21551,Coulter2017}. Observations were obtained between UT 00:08 and 00:20 on 2017 August 18. These images served both as confirmation of the transient \cite{Coulter2017} and acquisition images for the spectra of SSS17a \cite{GCN21547,Shappee2017}.

On subsequent nights, we continued to observe SSS17a with LDSS-3 in the \emph{Bgriz} bands. Observations were obtained on six additional nights spanning 2017 August 18 to 2017 August 25 and included $z$-band spectroscopic acquisition images and 2--5, 60~s exposures in the $B$- and $g$-bands, obtained once the transient had faded below the level accessible to the Swope telescope. In addition we observed SSS17a in the $r$-band on 2017 August 19 and the $z$-band on 2017 August 29 with the Inamori-Magellan Areal Camera and Spectrograph (IMACS; \cite{Dressler2006}) on the Magellan/Baade telescope.

Bias and flat field corrections were made to all LDSS-3 and IMACS images using standard routines in IRAF \cite{IRAF}. For observations obtained within 3.5~days of the LVC trigger, all exposures were reduced independently. After that date, nightly stacks were produced in each band. 

In order to account for host-galaxy light when performing photometry, we make use of the symmetry of the host galaxy. The galaxy light was subtracted after a 180 degree rotation around its center, thus flattening the target's background. Tests on several images using Pan-STARRS1 3$\pi$ images\cite{Chambers2016} of the field as templates yield consistent photometry, but worse residuals near the galaxy core. After galaxy subtraction, point-spread function (PSF) photometry was performed on each image using IRAF {\tt daophot} \cite{Stetson1987} tasks. Absolute calibration was performed using magnitudes of field stars. For observations in \emph{Bgri}, we use field star magnitudes calibrated by the Swope telescope. For $z$-band observations, our magnitudes are tied directly to Pan-STARRS1 magnitudes\cite{Flewelling2016}.  All magnitudes are listed in Table S1.

\subsection{Magellan/FourStar and du Pont/RetroCam near-IR Photometry}

We observed SSS17a using the FourStar near-IR camera \cite{Persson2013} mounted on the 6.5-m Magellan/Baade telescope and the RetroCam near-IR camera mounted on the 2.5-m du~Pont telescope at Las Campanas Observatory. FourStar observations were obtained on 11 nights beginning 11.5 hours after the LVC trigger as part of a joint Swope-Magellan search \cite{Coulter2017}. The observations span 2017 August 17 to 2017 September 7 (0.5 to 18.5~days post-LVC trigger) and were obtained in the $J1$ (approximately $Y$), $J$, $H$, and $K_{\rm{s}}$ filters. RetroCam observations were carried out over 6 nights between 2017 August 21 and 2017 August 27 (4.5 to 10.5 days post-LVC trigger) in $Y$, $J$ and $H$.    
Initial FourStar and Retrocam data reduction was performed following normal procedures of dark subtraction and flat fielding. In addition, a linearity correction was applied to FourStar data and a fringing mask was subtracted from RetroCam images. Dithered pointing sequences were then co-added into final image combinations, masking bad pixels.

SSS17a remained bright in the near-IR for over three weeks after the LVC trigger, during which the target was only visible close to twilight. It was therefore not possible to obtain high-quality near-IR template images for image subtraction. In order to account for host galaxy light from NGC\,4993 when measuring photometry of SSS17a, we again make use of the host galaxy symmetry, subtracting a 180 degree rotated image, as described above.

The instrumental PSF photometry was measured in most images using IRAF's {\tt daophot} package \cite{Stetson1987}. For some images with a low signal-to-noise ratio, an empirical PSF was measured on a bright star and fit via a Markov Chain Monte Carlo (MCMC) to the target and other stars in the field. In this case, a bright star in the image is selected, sky subtracted, subsampled, and then stored in a 2-dimensional array. This array is used as a model of the PSF, which is then fitted to the target and stars individually, using an MCMC procedure where the fitted variables are the amplitude and the center of the star. 

The final photometry of the target is measured relative to the calibrated stars in the field. The absolute calibration of field stars was made on one photometric night (2017 August 21) on du~Pont/RetroCam for the $Y$, $J$ and $H$ bands. $K_{\rm{s}}$-band calibration of field stars was tied to the Two Micron All Sky Survey (2MASS) catalog \cite{Skrutskie2006}.  Calibrations from 2MASS or du~Pont/Retrocam are fully consistent for the $J$ and $H$ bands.  Final photometry is presented in Table S1 on the AB magnitude system. Instrumental magnitudes were converted to the AB scale using offsets from \cite{Blanton2005} for $K_{\rm{s}}$ and \cite{Krisciunas17} for $Y$, $J$, and $H$.

The FourStar imaging of SSS17a obtained on the night of discovery used integration times long enough that the sky emission was saturated by the end of the exposure.  However, the data saved by the detector electronics consist of the difference between the voltage of each pixel at the end of the exposure and the voltage after the 1.456~s it takes for an initial read of each pixel at the beginning of the exposure.  A saturated image therefore contains a negative image of the field with an exposure time of 1.456~s. We reduced these data by subtracting a template of the saturation pattern and then taking the difference between consecutive dithered frames to remove any remaining sky emission or dark current.  We combined the available dithered frames to construct a final image. We then made photometric measurements as described above.

\subsection{Swift Ultraviolet and Optical Photometry}

We performed photometry on 9 epochs of publicly available {\it Swift} Ultraviolet/Optical Telescope (UVOT) observations of SSS17a taken between 2017 August 18 and 2017 August 21 \cite{GCN21550,GCN21572}. The UVOT observations were obtained in 6 filters: $v$ (effective wavelength 5402~\AA), $b$ (4328~\AA), $u$ (3494~\AA), $uvw1$ (2589~\AA), $uvm2$ (2228~\AA), and $uvw2$ (2030~\AA) \cite{Poole2008}. See S2.3 for discussion of the influence of the Swift-UVOT transmission functions on the analysis presented.

We extracted source counts from a 3.0\arcsec\ radius region around the transient using the software task {\sc Uvotsource}. In order to account for host galaxy light, we extracted background counts from several 3.0\arcsec\ radius regions at a similar distance from the host galaxy core. Count rates were converted into magnitudes and fluxes using the most recent UVOT calibrations \cite{Poole2008,Breeveld2010}.

Our results produce UV and $u$-band magnitudes that are $\sim$0.1$-$0.2 mag fainter than the values reported in \cite{GCN21550} and \cite{GCN21572}. The deviation is larger in redder bands, where the host galaxy is brighter, and we attribute this difference to our method of background subtraction. If we instead choose a background region that is well separated from NGC 4993, we recover magnitudes consistent with those previously reported in the Gamma-ray Coordinates Network (GCN) circulars \cite{GCN21550,GCN21572}. Applying the same methods described above to the {\it Swift} $b$ and $v$ band images shows evidence for continued galaxy contamination, in the form of a flat light curve with large uncertainties over a $\sim$2 day timespan, which conflicts with evolution seen in higher resolution ground-based (e.g LDSS-3) imaging at similar wavelengths. In this manuscript, we report only values from {\it Swift} $u$, $uvw1$, $uvm2$ and $uvw2$ bands. All measured values are reported in Table S1 on the AB scale.

\subsection{PESSTO EFOSC and SOFI Imaging}

The Public ESO Spectroscopic Survey of Transient Objects (PESSTO) \cite{Smartt2015} obtained optical imaging of SSS17a with the ESO Faint Object Spectrograph and Camera v.2 (EFOSC2) \cite{Buzzoni1984} and near-IR imaging with Son OF ISAAC (SOFI) \cite{Moorwood1998} on the New Technology Telescope (NTT) beginning on 2017 August 18 \cite{GCN21582}.  We downloaded the raw PESSTO observations from the ESO archive and reduced them using the Pyraf pipeline described by \cite{Smartt2015}.  The PESSTO optical imaging consists of three 300~s exposures in $U$ and several $10-60$~s exposures in $V$ per night on the nights of 2017 August 18-19, 2017 August 19-20, and 2017 August 20-21.  The PESSTO near-IR imaging consists of a sequence of $48-67$ dithered 90~s or 120~s exposures in a single band ($J$, $H$, or $K_{\rm s}$) per night.

We performed PSF photometry on the $U$- and $V$-band images using the {\tt IRAF} task {\tt daophot} and calibrated the instrumental magnitudes using stars in the Swope $V$- and \swift\ $U$-band images.  Photometry on the SOFI $J$, $H$, and $K_{\rm s}$ images was carried out as described above for Magellan/FourStar.

\subsection{Keck Imaging}

We imaged SSS17a with the Low Resolution Imaging Spectrometer (LRIS) \cite{Oke1995,Steidel2004} on the Keck-I 10~m telescope on 24 August 2017 from UT 05:35 to 06:24.  LRIS was configured in imaging mode with $g$ and $I$ filters on the blue and red sides, respectively, and the D560 dichroic.  We observed SSS17a in $11$ blue-side frames and $27$ red-side frames starting during twilight, and our exposure times varied from $14$--$124$~s on the blue side and $25$--$120$~s on the red side.  In order to minimize readout time on the red side, we binned the image $2 \times 2$ and read out only the central $4\mbox{~arcmin} \times 4\mbox{~arcmin}$ region.  Conditions were near photometric at the start of observations with $\sim$0.8\arcsec\ seeing, although the seeing degraded significantly in exposures at higher airmasses.  We obtained bias frames and sky flats in the same instrumental configuration and the Landolt standard star field SA95 275 \cite{Landolt1992} was observed at a similar airmass.

We reduced all Keck/LRIS images using the semi-automatic IDL/python software {\tt LPipe} \cite{LPipe}.  Individual frames were bias corrected, flattened, and then registered using stars in the 2MASS Point Source Catalog \cite{Skrutskie2006}.  We aligned and co-added the registered images, weighting by the inverse variance of the emission-free regions.  We performed PSF photometry using {\tt daophot} \cite{Stetson1987} and calibrated our photometry using APASS and Landolt standard stars \cite{Landolt1992,Henden2016} from our standard star field.

\subsection{Synthetic Photometry from Flux-Calibrated Spectra}

In addition, we supplement the photometry measured from broad-band imaging described above with synthetic photometry performed on flux calibrated spectra of SSS17a. Full details of the spectroscopic observations, reductions, calibration and synthetic photometry are described in \cite{Shappee2017}.  Briefly, the final flux calibrated spectra were corrected for instrumental response, atmospheric dispersion, differential flux losses, and telluric absorption, and were then scaled to the broad-band photometry described above. 

The full set of synthetic photometry measured from the spectral sequence of SSS17a between 0.5 and 4.5 days can be found in \cite{Shappee2017}.  In this manuscript, we supplement our existing light curves and SEDs with measurements from synthetic photometry under the following conditions: (1) they add to a band in which we had existing observations, (2) they were obtained from a high S/N spectrum, (3) the entire filter response function falls within the observed spectrum, and (4) the relevant band lies interior or adjacent to the broad-band photometric points used to scale the spectrum.  This results in the addition of {\it V}, {\it r}, and {\it z}-band measurements from the first night of observations and individual {\it V} and {\it z}-band points on later epochs. Synthetic photometry used in this analysis is presented in Table S1.

\section{Photometric Analysis}

\subsection{Distance and Reddening}

SSS17a exploded in NGC\,4993 at a distance of $\sim$40 Mpc. Throughout this manuscript we adopt the Tully-Fisher distance of 39.5 Mpc from \cite{Freedman2001}.  

We adopt a value of $E(B-V)$ = 0.106 for the color excess due to Milky Way reddening in the direction of SSS17a based on the dust maps of \cite{Schlafly2011}. $E(B-V) \equiv A_{B} - A_{V}$ where $A_{B}$ and $A_{V}$ are the total extinction in the $B$ and $V$ band, respectively. This value is consistent with estimates based on Milky Way Na D absorption lines observed in high resolution spectra of SSS17a\cite{Shappee2017}. Photometry is corrected for this reddening using a Milky Way extinction curve from \cite{Cardelli1989} with the parameter $R \equiv$ $A_{{V}}$/$E(B-V)$ $=$ 3.1. We use the SED of the source to iteratively calculate the extinction in each filter, A$_{\lambda}$, at each epoch as described below. Throughout this manuscript, we do not correct for reddening due to dust within the host galaxy. The lack of observed Na D features at the redshift of NGC4993 favors a low intrinsic absorption \cite{Shappee2017}. 

\subsection{Basic Photometric Properties from the UV to near-IR}

The full light curve of SSS17a from the UV through near-IR is shown in Figure 1. In Table S2 we provide several basic parameters derived from the photometry in each band.

The quoted peak apparent and absolute magnitudes are observed values. For bands blueward of z-band, which rise in less than 12 hours, these are lower limits to the true peak. SSS17a peaks between $-$15.5 and $-$16.0 mag (AB) from the optical to near-IR, similar to the faintest core-collapse supernovae. In Figure S1 we plot the {\it V}, {\it i}, and {\it H}-band absolute magnitude light curves of SSS17a---accounting for the distance to NGC4993 and Milky Way reddening. Also shown are the F160W (roughly rest-frame {\it H}-band) measurement and F6060W (roughly rest-frame {\it V}-band) upper limit for the tentative kilonova associated with short GRB 130603B \cite{Berger2013,Tanvir2013}. At $+$7 days from the burst, GRB130603B has a slightly higher near-IR luminosity than observed for SSS17a. 

In Table S2 we also report several measurements of the rise and decline timescales of SSS17a. We determine the time over which a band declines by half of its peak luminosity (t$_{1/2}$) and number of magnitudes the light curve decline in the first 5 days after maximum ($\Delta$M$_{5}$) by linearly interpolating the observed light curves. A linear decline rate was also obtained by fitting the data between maximum light and +5.5 days. SSS17a declines at a rate ranging from 2.6 mag day$^{-1}$ ($u$-band) to 0.1 mag day$^{-1}$ ($H$-band). The observed optical light curves are generally consistent with a single slope post-maximum. However, our final $B$-band points indicate a shallowing of the decay rate between 5 and 10 days post-merger. 

SSS17a peaks earlier and declines faster in bluer bands. As a result, its colors transition quickly to the red. Between +0.5 and +4.5 days the $V-H$ color of SSS17a evolves from $-$1.2 mag to 3.6 mag (top panel, Figure S1). By comparison, the late time emission from GRB130603B was attributed to a kilonova in part because of its red color ($V-H$ $>$ 1.9 mag). 

The characteristic timescale of SSS17a is faster than other known transients. For example, t$_{1/2}$ $\sim$ 1.1 days in r-band is nearly a factor of 5 smaller than the fastest rapidly-evolving and luminous transient discussed in \cite{Drout2014}. A detailed comparison of the properties of SSS17a to other transients is made in \cite{Siebert17}.

\subsection{SED and Bolometric Light Curve Construction}

We construct SEDs for SSS17a from our observed photometry at 10 epochs using the iterative, forward-modeling approach described in \cite{Brown2016}. This approach aims to mitigate systematic errors that occur when constructing an SED from photometry by applying standard offsets to obtain fluxes at filter's pivot or effective wavelength. The translation from an observed broad-band magnitude to a flux at a given wavelength is fundamentally a function of the source's spectrum, and systematic errors can be particularly large when applying standard conversion factors to Swift-UVOT $uvw1$ and $uvw2$ observations of red sources \cite{Brown2016}.

We begin by interpolating our observed light curves to a set of common epochs. Our observations predominately come from Las Campanas Observatory and other telescopes in Chile, where SSS17a was only visible for the first $\sim$1 hour of the night. As a result our SEDs are constructed predominately on daily, evenly-spaced epochs between 0.5 and 8.5 days after the LVC trigger. In addition, we also compute compute SEDs at the 0.67 and 1.0 days. These correspond to the first two epochs of Swift UVOT data, when the transient was rapidly evolving.

With our photometry integrated to common epochs, we then construct a simple SED with ``pivot'' points at the effective wavelength of each observed band and the outer edge of the filter functions for the exterior bands. The flux level at each pivot point is then adjusted until synthetic photometry performed on this simple SED is consistent with the observed measurements. Extinction correction factors, A$_{\lambda}$, and flux conversion factors for each band are then computed from this best-fitting SED and applied to the original data.  

The time evolution of the resulting SEDs is shown in Figure 3.  In Figure S2 we plot each epoch individually for clarity. As in Figure 3, the shaded bands represent best-fit blackbody temperatures for each epoch.  For epochs 0.67 and 1.0 days, data redward of the Swift-u band data is derived from interpolating our ground based optical data on days 0.5 and 1.5. We find very similar best-fit blackbody temperatures if only the Swift-UVOT data are modeled. 

SSS17a undergoes dramatic cooling over the first day post-merger. Between our first observations at 0.5 days, and the first epoch of {\it Swift-}UVOT measurements $\sim$ 4 hours later, we infer a total cooling of $\sim$2400K, or 600 K hr$^{-1}$. Evidence for this rapid cooling is also seen in the evolution of the continuum between spectra taken $\sim$ 1 hour apart on the first night of observations \cite{Shappee2017}.

While the SEDs are generally consistent with a thermal distribution, there are a few exceptions. Between days 1.5 and 4.5 the Y-band measurements consistently lead to fluxes in excess of the best-fit blackbody distribution. This is consistent with the emergence of a feature in in the optical spectra around $\sim$ 1$\mu$m at similar epochs\cite{Shappee2017}. In addition, an apparent excess is visible in the $g$-band ($\sim$4800 \AA\,) at similar epochs. These features are indicated by dotted lines in Figure S2. At late epochs ($>$ 5.5 days) the slope between the bluest observed bands are steeper than a thermal distribution, possibly indicative of line-blanketing. 

Pseudo-bolometric luminosities at each epoch were constructed by integrating these best-fit SEDs after applying a Milky Way extinction curve, and corrected for missing flux as described in the main text. At 0.5 days post-merger SSS17a has a bolometric flux of $\sim$10$^{42}$ erg s$^{-1}$.

\section{R-Process Heating and Ejecta Property Estimates}

Models of r-process powered transients have become increasingly sophisticated over the last decade. Early models first showed that radioactive ejecta from a neutron star merger can power an electromagnetic transient, but assumed that the energy was provided by some fraction of the rest mass energy rather than using an actual model for the radioactive decay \cite{Li1998}. The idea of a macronova or kilonova was then re-popularized by newer models that invoked the decay of radioactive species, without making the connection to r-process nucleosynthesis \cite{Kulkarni2005,Metzger2008}. The first true r-process powered transient calculation soon followed \cite{Metzger2010}, and since that time the theoretical studies have quickly matured with more detailed nuclear networks, hydrodynamics in multi-dimensional numerical simulations, and detailed treatments of the radiative transfer. Here we summarize some of the basic aspects of r-process transients for use in comparing with our data. A more detailed comparison with specific models is provided in a companion paper \cite{Kilpatrick17}.

One of the most robust predictions of r-process heating is the $t^{-1.3}$ dependence of the energy generation rate, $\dot{q}_r$  \cite{Metzger2010,Roberts2011}. We fit the heating rate from these works to find 
\begin{equation}
    \dot{q}_r = 3\times10^{10}t_{\rm day}^{-1.3}\,{\rm erg\,s^{-1}\,g^{-1}}, 
\end{equation}
where $t_{\rm day}$ is the time since explosion in units of days. There can be a $\approx$30\% correction to this heating rate depending on $Y_e$. The thermalization depends on the ability to absorb the energy with energy losses from weak processes. To account for thermalization efficiency, $\epsilon_{\rm th}$, we use the fitting function
\begin{equation}
    \epsilon_{\rm th} = 0.36 \left[ 
    \exp(-a t_{\rm day})
    +\frac{\left( 1+2bt^d\right)}{2bt_{\rm day}^d}
    \right],
\end{equation}
where $a$, $b$, and $d$ are fitting parameters whose values depend on the mass of of r-process elements, $M_{\rm{r}}$ and the average (lowest) velocity in the ejecta, $v_0$. $a=0.56$, $b=0.17$, and $d=0.74$ are appropriate for $M_{\rm{r}}=0.1\,M_\odot$ and $v_0=0.1c$ \cite{Barnes2016}. Combining the above expressions the total r-process luminosity is
\begin{equation}
    L_r = \epsilon_{\rm th}\dot{q}_r M_{r},
\end{equation}

The rise time to peak ($t_p$) for an optically thick transient is roughly set by when the diffusion time is equal to the expansion time for the ejecta \cite{Arnett1982}. For ejecta with a velocity gradient $v= v_0(M_v/M_{r})^{-1/\beta}$, where $M_v$ is the total mass with velocity greater than $v$, this timescale can be approximated as
\begin{equation}
    t_p \approx \left( \frac{\kappa M_r}{3\beta v_0 c} \right)^{1/2},
\end{equation}
where $\kappa$ is the specific opacity \cite{Metzger2017}. A characteristic value for $\beta$ motivated by numerical simulations is about 3 \cite{Bauswein2013,Rosswog2014}, although in general the ejecta structure can be more complicated than in such simulations \cite{Piran2013}. Solving this equation for $\kappa$ and requiring that $t_p \lesssim 0.5\,{\rm days}$, we find an upper limit to the opacity during the peak of the kilonova of $\kappa\approx0.08\,{\rm cm^2\,g^{-1}}$.

Similarly, the optical depth of a given depth moving at velocity $v$ is
\begin{equation}
    \tau_v  = \int_r^{\infty} \kappa \rho_v dr,
\end{equation}
where
\begin{equation}
    \rho_v \approx \frac{3M_v}{4\pi (vt)^3}= \frac{3M_{r}}{4\pi (vt)^3}
    \left(\frac{v}{v_0} \right)^{-\beta}
\end{equation}
is roughly the density of the material moving at $v$. Performing this integral and evaluating it at $v=v_0$, we find that the total optical depth is
\begin{equation}
    \tau = \frac{3\kappa M_r}{4\pi (1+\beta)t^2v_0^2}.
\end{equation}
Solving for the time when $\tau\approx1$
\begin{equation}
    t_{\tau=1} \approx \left[\frac{3\kappa M_r}{4\pi(1+\beta)v_0^2} \right]^{1/2},
\end{equation}
thus up to a factor of order unity $t_{\tau=1}\approx(c/v_0)^{1/2}t_p$. Setting
$t_{\tau=1}\approx20\,{\rm days}$ (motivated by the observed duration of the SSS17a emission), we derive a lower limit on the opacity of the red/cool component of $\kappa\gtrsim5\,{\rm cm^2\,g^{-1}}$.

\section{Constraints on Radioactive Nickel Heating}

Most hydrogen-deficient supernovae are powered by the radioactive decay of $^{56}$Ni, but radioactive nickel heating cannot reproduce the entire photometric evolution of SSS17a. The heating from the decay of $^{56}$Ni and subsequently $^{56}$Co, $\dot{q}_{56}$,  scales as
\begin{equation}
    \dot{q}_{56} = \dot{q}_{\rm Ni}
    \exp(-t/\tau_{\rm Ni})
    + \dot{q}_{\rm Co}
    [\exp(-t/\tau_{\rm Co}) - \exp(-t/\tau_{\rm Ni})]
\end{equation}
where $\dot{q}_{\rm Ni} = 3.9\times10^{10}\,{\rm erg\,g^{-1}\,s^{-1}}$, $\dot{q}_{\rm Co} = 6.8\times10^{9}\,{\rm erg\,g^{-1}\,s^{-1}}$, $\tau_{\rm{Ni}} = 8.8\,{\rm days}$, and $\tau_{\rm Co}=111.3\,{\rm days}$, are the heating rates and decay timescales for $^{56}$Ni and $^{56}$Co. The total amount of heating depends on how efficiently gamma-rays are thermalized. Gamma-ray thermalization efficiency, $\epsilon_{56}$, can be simply modeled with the following function \cite{Wheeler15}
\begin{equation}
    \epsilon_{56} = 
        1 - \exp[-(T_0/t)^2],
\end{equation}
where the characteristic gamma-ray diffusion timescale is \cite{Clocchiatti97}
\begin{equation}
    T_0 = \left( \frac{10C\kappa_\gamma M_{\rm ej}}{3v^2}\right)^{1/2},
\end{equation}
where $\kappa_\gamma = 0.03\,{\rm cm^2\,g^{-1}}$ is the fiducial gamma-ray opacity, $M_{\rm ej}$ is the total ejecta mass, and $C\sim0.05$ is a structural factor that depends on the density distribution. A larger $T_0$ means that the gamma-rays take longer to diffuse and thus are more efficiently thermalized. The total luminosity is
\begin{equation}
    L_{56} = \epsilon_{56} \dot{q}_{56} M_{56},
\end{equation}
where $M_{56}$ is the total $^{56}$Ni mass.

In Figure S3, we compare the bolometric light curve of SSS17a to nickel curves with $M_{56}=0.014\,M_\odot$, the nickel mass required to match the first night's luminosity. Since the total ejecta mass can be different from the nickel mass, we consider a range of $T_0=1-10\,{\rm days}$ (and in addition the least shallow curve is for $T_0=\infty$, or complete thermalization of gamma-rays), corresponding to a range of ejecta masses from $M_{\rm ej}=0.002-0.2\,M_\odot$. The lowest value is unphysical since it is smaller than $M_{\rm  56}$, but we include it just to demonstrate that no nickel-decay model can reproduce both the peak luminosity and subsequent evolution of SSS17a. 
In Figure S3 we also plot a nickel curve with $M_{56}=0.002\,M_\odot$ and $T_0=10\,{\rm days}$, corresponding to an ejecta mass of $\sim$0.2 M$_\odot$ for $v = 0.1c$. This curve reproduces the bolometric evolution of SSS17a between 7.5 and 18.5 days post-merger, but requires that another emission source dominates at early times. In addition, a transient powered by the radioactive decay of nickel does not naturally explain the color evolution observed in SSS17a.

\clearpage

\begin{figure}
\begin{center}
\includegraphics*[width=0.9\textwidth,trim=0 3cm 0 4cm, clip]{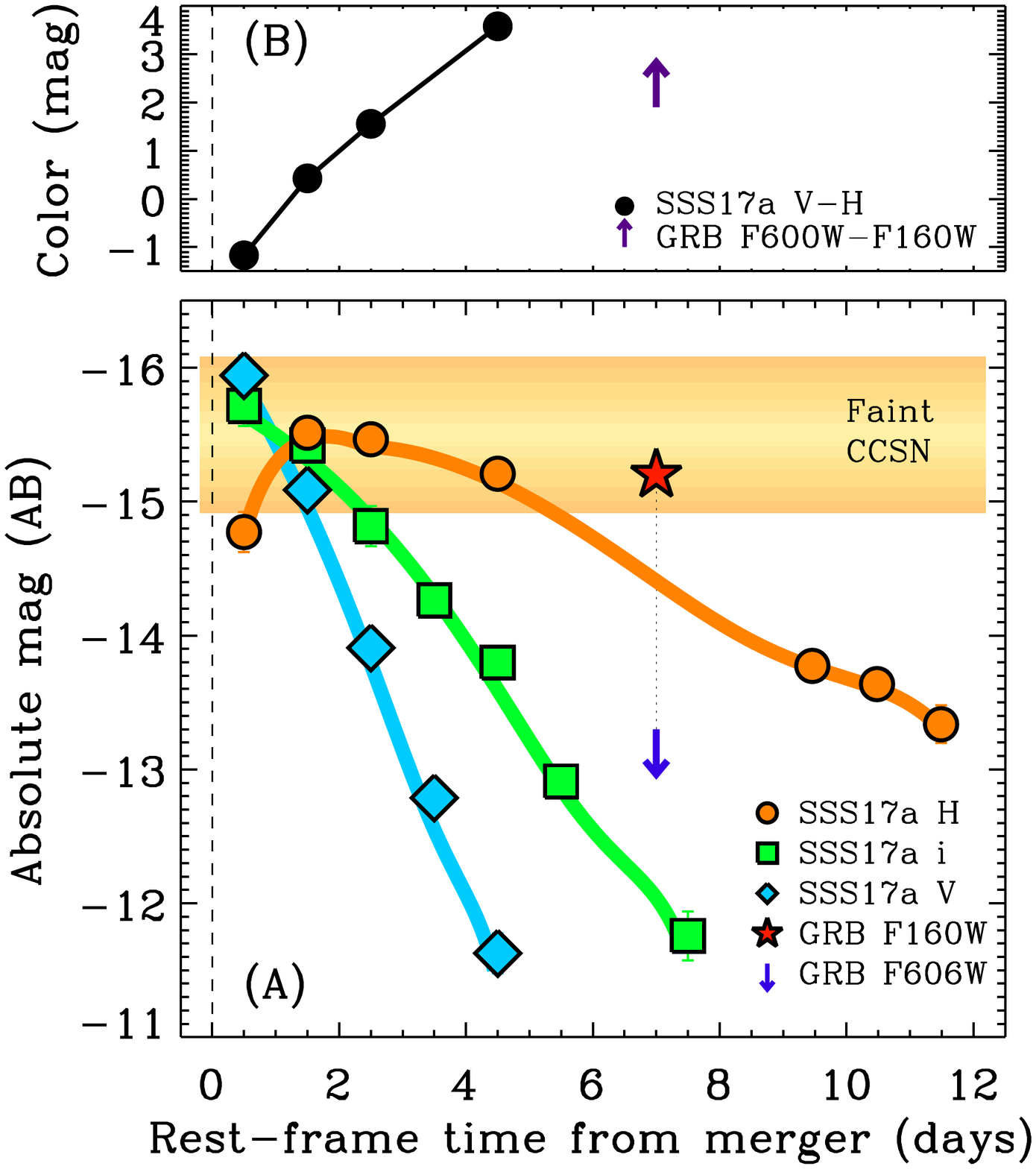}
\caption{{\bf Absolute magnitude light curves and $V-H$ color curves of SSS17a and the tentative kilonova associated with GRB130603B.} Vertical dashed lines indicate the time of the merger. {\bf (A):} Absolute magnitude light curves of SSS17a in the {\it H}, {\it i}, and {\it V}-bands (orange circles, green squares and cyan diamonds, respectively). SSS17a reaches similar peak absolute magnitudes to faint core-collapse SNe (shaded band). Also shown are observations of the tentative kilonova  associated with GRB130603B from \cite{Berger2013,Tanvir2013}. These data correspond to roughly rest-frame {\it H}-band (red star) and {\it V}-band (blue arrow; upper limit). At $+$7 rest frame days, GRB130603B has a higher near-IR luminosity than SSS17a. {\bf (B):} $V-H$ color. SSS17a undergoes rapid reddening in the first five days post merger. A lower limit on the color of GRB130603B is indicated by the purple arrow.}
\end{center}
\end{figure}

\begin{figure}
\begin{center}
\includegraphics*[width=\textwidth]{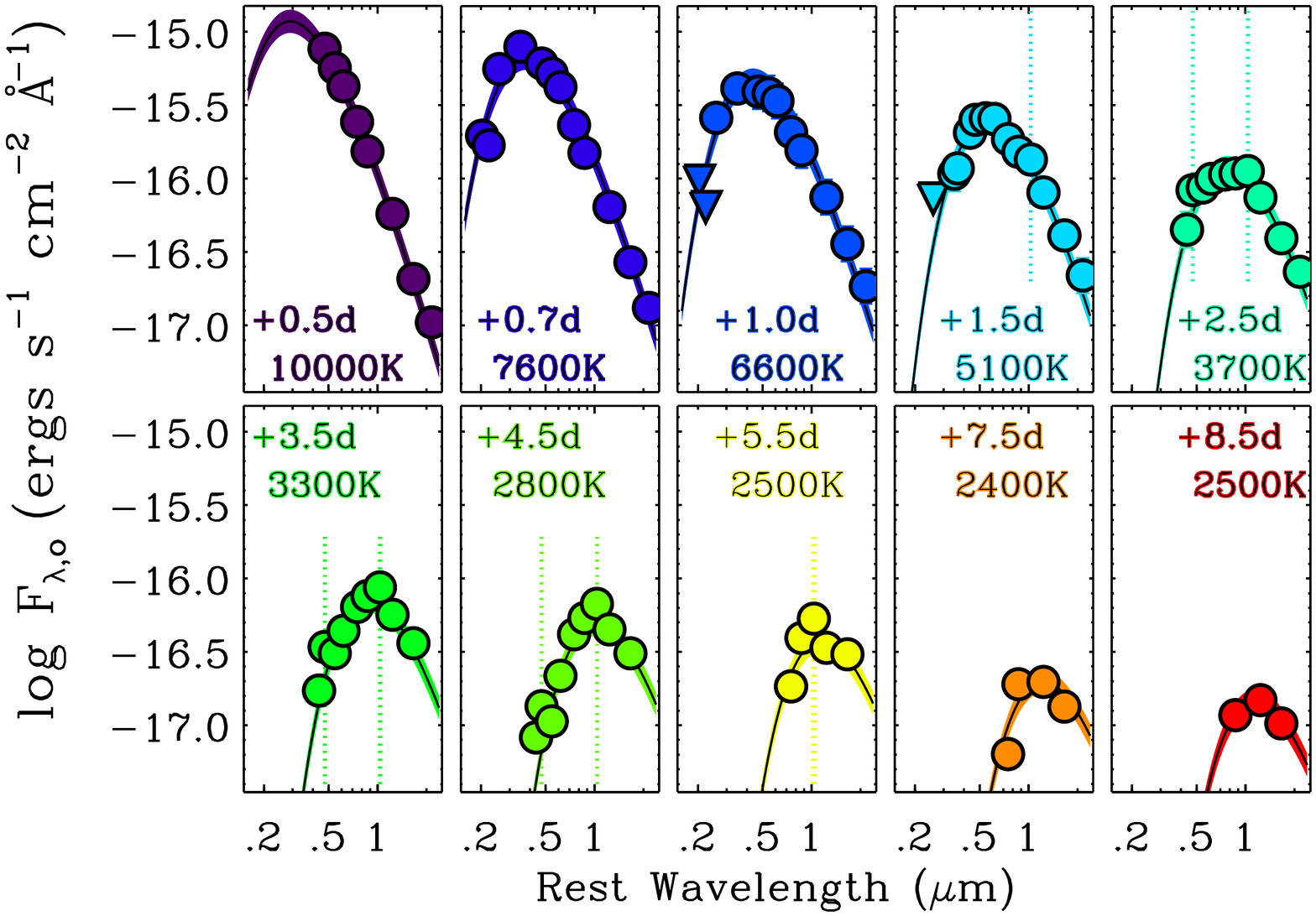}
\caption{{\bf Evolution of the spectral energy distribution of SSS17a, shown separately for each epoch.} Flux units are the same as in Figure 3. Detections as plotted as circles and upper limits as downward pointing triangles. Solid lines represent best-fitting blackbody model distributions. SSS17a underwent a rapid cooling in the first 48 hours post-merger. At later times the best-fit color temperature asymptotically approaches $\sim$2500 K.}
\end{center}
\end{figure}

\begin{figure}
\begin{center}
\includegraphics*[width=1.0\textwidth,trim=0 3cm 0 10cm, clip]{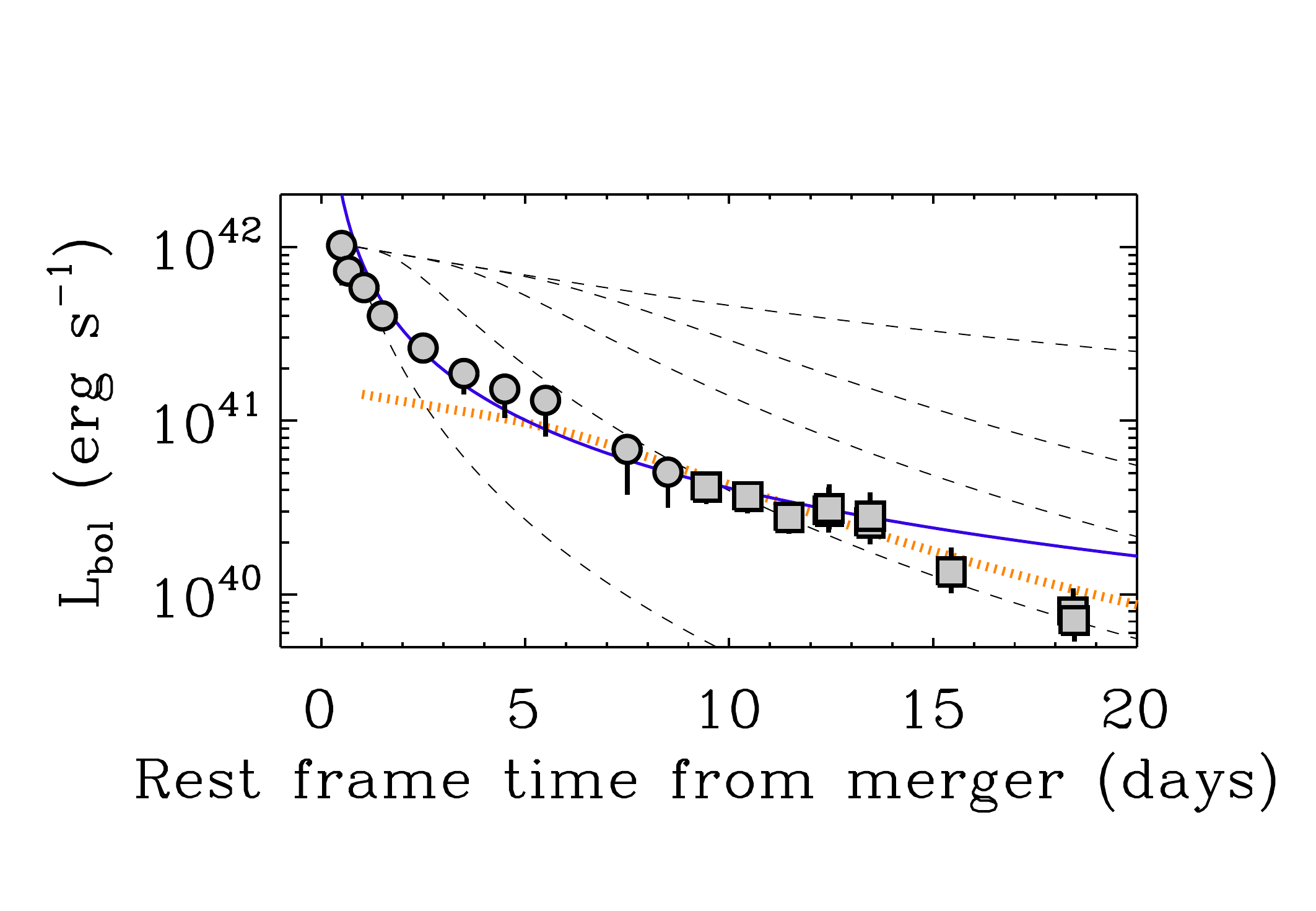}
\caption{{\bf Radioactive nickel and r-process heating in comparison to the bolometric light curve of SSS17a.} Data is the same as in Figure 4A. Thin black dashed lines are the heating rate from $^{56}$Ni with varying levels of gamma-ray leakage with $T_0=\infty, 10, 6, 3,$ and $1\,$day (from top to bottom). None of these curves have the correct shape to match SSS17a. In comparison, the r-process heating (blue solid curve) better matches the data at least out to $\sim$13 days. A nickel curve with a total nickel mass of 0.002 M$_\odot$ and total ejecta mass of $\sim$0.2 M$_\odot$ (orange dotted line) can reproduce the bolometric evolution from day 7.5 to 18.5 if another emission source dominates at early times, but does not naturally explain the late-time color evolution observed for SSS17a.}
\end{center}
\end{figure}

\clearpage

\begin{longtable}{@{}lcccr}
\caption{{\bf UV to near-IR photometry of SSS17a.} Columns include the Julian Date of the observations (JD), rest frame days since the LVC trigger (phase), the broad-band filter of the observation, the observed magnitude of SSS17a (m$_{\rm{obs}}$), and the telescope/instrument on which the observation was acquired. All magnitudes are on the AB scale and have not been corrected for foreground Milky Way reddening. Magnitude errors are given in parenthesis next to the observed magnitude and upper limits are designated with the symbol, ``$>$''. Measurements derived from performing synthetic photometry on flux calibrated spectra are designated with a ``-s'' in the facility/instrument column. This table is also provided in machine readable form.}\\
\hline
\hline
JD & Phase & Filter & m$_{\rm{obs}}$ ($\sigma$) & Facility/Instrument\\ 
 & (day) & & (AB mag) & \\\hline
2457983.6690   &   0.67   &   \textit{uvw2}   &   21.11 (0.22)   &   Swift-UVOT\\
2457984.0730   &   1.07   &   \textit{uvw2}   &   $>$ 21.87   &   Swift-UVOT\\
2457984.5330   &   1.53   &   \textit{uvw2}   &   $>$ 21.84   &   Swift-UVOT\\
2457985.0640   &   2.06   &   \textit{uvw2}   &   $>$ 22.49   &   Swift-UVOT\\
2457985.3200   &   2.32   &   \textit{uvw2}   &   $>$ 22.61   &   Swift-UVOT\\
2457985.6520   &   2.65   &   \textit{uvw2}   &   $>$ 22.45   &   Swift-UVOT\\
2457985.8620   &   2.86   &   \textit{uvw2}   &   $>$ 23.13   &   Swift-UVOT\\
2457986.5820   &   3.58   &   \textit{uvw2}   &   $>$ 22.14   &   Swift-UVOT\\
2457983.6520   &   0.65   &   \textit{uvm2}   &   21.13 (0.22)   &   Swift-UVOT\\
2457984.0640   &   1.06   &   \textit{uvm2}   &   $>$ 22.18   &   Swift-UVOT\\
2457984.5150   &   1.51   &   \textit{uvm2}   &   $>$ 22.34   &   Swift-UVOT\\
2457985.0680   &   2.07   &   \textit{uvm2}   &   $>$ 22.16   &   Swift-UVOT\\
2457985.4640   &   2.46   &   \textit{uvm2}   &   $>$ 20.49   &   Swift-UVOT\\
2457985.6630   &   2.66   &   \textit{uvm2}   &   $>$ 22.50   &   Swift-UVOT\\
2457985.9290   &   2.93   &   \textit{uvm2}   &   $>$ 23.07   &   Swift-UVOT\\
2457986.5930   &   3.59   &   \textit{uvm2}   &   $>$ 22.91   &   Swift-UVOT\\
2457983.6610   &   0.66   &   \textit{uvw1}   &   19.33 (0.10)   &   Swift-UVOT\\
2457984.0680   &   1.07   &   \textit{uvw1}   &   20.20 (0.22)   &   Swift-UVOT\\
2457984.5240   &   1.52   &   \textit{uvw1}   &   $>$ 21.54   &   Swift-UVOT\\
2457985.0720   &   2.07   &   \textit{uvw1}   &   $>$ 21.73   &   Swift-UVOT\\
2457985.2730   &   2.27   &   \textit{uvw1}   &   $>$ 21.67   &   Swift-UVOT\\
2457985.6440   &   2.64   &   \textit{uvw1}   &   $>$ 21.36   &   Swift-UVOT\\
2457985.8540   &   2.85   &   \textit{uvw1}   &   $>$ 22.06   &   Swift-UVOT\\
2457986.5750   &   3.58   &   \textit{uvw1}   &   $>$ 21.85   &   Swift-UVOT\\
2457983.6670   &   0.67   &   \textit{u}   &   18.16 (0.07)   &   Swift-UVOT\\
2457984.0100   &   1.01   &   \textit{u}   &   18.87 (0.15)   &   Swift-UVOT\\
2457984.0720   &   1.07   &   \textit{u}   &   19.00 (0.15)   &   Swift-UVOT\\
2457984.5300   &   1.53   &   \textit{u}   &   20.32 (0.30)   &   Swift-UVOT\\
2457985.0600   &   2.06   &   \textit{u}   &   $>$ 21.46    &   Swift-UVOT\\
2457985.3160   &   2.32   &   \textit{u}   &   20.63 (0.32)   &   Swift-UVOT\\
2457985.6480   &   2.65   &   \textit{u}   &   $>$ 20.55   &   Swift-UVOT\\
2457985.8580   &   2.86   &   \textit{u}   &   $>$ 21.30   &   Swift-UVOT\\
2457986.5780   &   3.58   &   \textit{u}   &   $>$ 20.94   &   Swift-UVOT\\
2457984.5481   &   1.55   &   \textit{U}   &   20.11 (0.23)   &   NTT/EFOSC2\\
2457984.5520   &   1.55   &   \textit{U}   &   20.21 (0.28)   &   NTT/EFOSC2\\
2457984.5559   &   1.56   &   \textit{U}   &   20.10 (0.28)   &   NTT/EFOSC2\\
2457985.5537   &   2.55   &   \textit{U}   &   $>$ 20.19   &   NTT/EFOSC2\\
2457986.5569   &   3.56   &   \textit{U}   &   $>$ 20.49   &   NTT/EFOSC2\\
2457984.4829   &   1.48   &   \textit{B}   &   19.04 (0.06)   &   Magellan/LDSS\\
2457984.4843   &   1.48   &   \textit{B}   &   19.04 (0.07)   &   Magellan/LDSS\\
2457984.5342   &   1.53   &   \textit{B}   &   19.07 (0.04)   &   Swope\\
2457986.4951   &   3.50   &   \textit{B}   &   21.72 (0.13)   &   Swope\\
2457987.5217   &   4.52   &   \textit{B}   &   22.52 (0.14)   &   Magellan/LDSS\\
2457990.5302   &   7.53   &   \textit{B}   &   23.85 (0.31)   &   Magellan/LDSS\\
2457983.5057   &   0.51   &   \textit{g}   &   17.41 (0.02)   &   Magellan/LDSS\\
2457983.5106   &   0.51   &   \textit{g}   &   17.41 (0.04)   &   Magellan/LDSS\\
2457983.5141   &   0.51   &   \textit{g}   &   17.39 (0.02)   &   Magellan/LDSS\\
2457984.4814   &   1.48   &   \textit{g}   &   18.61 (0.03)   &   Magellan/LDSS\\
2457984.4875   &   1.49   &   \textit{g}   &   18.66 (0.03)   &   Magellan/LDSS\\
2457984.5548   &   1.55   &   \textit{g}   &   18.49 (0.12)   &   Swope\\
2457986.5013   &   3.50   &   \textit{g}   &   20.77 (0.05)   &   Swope\\
2457987.5044   &   4.50   &   \textit{g}   &   21.75 (0.10)   &   Swope\\
2457987.5192   &   4.52   &   \textit{g}   &   21.78 (0.06)   &   Magellan/LDSS\\
2457989.7347   &   6.73   &   \textit{g}   &   $>$ 22.20   &   Keck/LRIS\\
2457990.5361   &   7.54   &   \textit{g}   &   $>$ 22.58   &   Magellan/LDSS\\
2457991.5340   &   8.53   &   \textit{g}   &   $>$ 22.64   &   Magellan/LDSS\\
2457983.5000   &   0.50   &   \textit{V}   &   17.35 (0.02)   &   Magellna/LDSS-s\\
2457984.4640   &   1.46   &   \textit{V}   &   18.22 (0.08)   &   NTT/EFOSC2\\
2457984.4687   &   1.47   &   \textit{V}   &   18.16 (0.05)   &   NTT/EFOSC2\\
2457984.4701   &   1.47   &   \textit{V}   &   18.13 (0.08)   &   NTT/EFOSC2\\
2457984.4948   &   1.49   &   \textit{V}   &   18.22 (0.04)   &   Swope\\
2457984.5105   &   1.51   &   \textit{V}   &   18.14 (0.04)   &   NTT/EFOSC2\\
2457984.5118   &   1.51   &   \textit{V}   &   18.16 (0.06)   &   NTT/EFOSC2\\
2457984.5129   &   1.51   &   \textit{V}   &   18.18 (0.04)   &   NTT/EFOSC2\\
2457984.5461   &   1.55   &   \textit{V}   &   18.25 (0.06)   &   NTT/EFOSC2\\
2457984.5471   &   1.55   &   \textit{V}   &   18.18 (0.10)   &   NTT/EFOSC2\\
2457985.4714   &   2.47   &   \textit{V}   &   19.40 (0.11)   &   NTT/EFOSC2\\
2457985.5000   &   2.50   &   \textit{V}   &   19.51 (0.08)   &   Magellan/IMACS-s\\
2457985.5164   &   2.52   &   \textit{V}   &   19.53 (0.12)   &   NTT/EFOSC2\\
2457985.5537   &   2.55   &   \textit{V}   &   19.59 (0.20)   &   NTT/EFOSC2\\
2457986.4687   &   3.47   &   \textit{V}   &   20.54 (0.20)   &   NTT/EFOSC2\\
2457986.4893   &   3.49   &   \textit{V}   &   20.52 (0.12)   &   Swope\\
2457986.5156   &   3.52   &   \textit{V}   &   20.55 (0.15)   &   NTT/EFOSC2\\
2457986.5531   &   3.55   &   \textit{V}   &   20.68 (0.31)   &   NTT/EFOSC2\\
2457987.5000   &   4.50   &   \textit{V}   &   21.85 (0.22)   &   Magellan/LDSS-s\\
2457983.5000   &   0.50   &   \textit{r}   &   17.33 (0.02)   &   Magellan/LDSS-s\\
2457984.4806   &   1.48   &   \textit{r}   &   17.91 (0.03)   &   Magellan/LDSS\\
2457984.5472   &   1.55   &   \textit{r}   &   17.98 (0.02)   &   Swope\\
2457985.5094   &   2.51   &   \textit{r}   &   18.93 (0.02)   &   Magellan/IMACS\\
2457986.5076   &   3.51   &   \textit{r}   &   19.82 (0.09)   &   Swope\\
2457987.4965   &   4.50   &   \textit{r}   &   20.58 (0.12)   &   Swope\\
2457988.4645   &   5.46   &   \textit{r}   &   $>$ 19.84    &   Magellan/IMACS\\
2457983.4814   &   0.48   &   \textit{i}   &   17.48 (0.02)   &   Swope\\
2457984.4798   &   1.48   &   \textit{i}   &   17.77 (0.03)   &   Magellan/LDSS\\
2457984.5438   &   1.54   &   \textit{i}   &   17.80 (0.02)   &   Swope\\
2457985.5000   &   2.50   &   \textit{i}   &   18.36 (0.02)   &   Magellan/IMACS-s\\
2457986.5046   &   3.50   &   \textit{i}   &   18.92 (0.05)   &   Swope\\
2457987.4885   &   4.49   &   \textit{i}   &   19.39 (0.04)   &   Swope\\
2457988.5020   &   5.50   &   \textit{i}   &   20.27 (0.12)   &   Swope\\
2457990.4897   &   7.49   &   \textit{i}   &   21.42 (0.18)   &   Swope\\
2457989.7344   &   6.73   &   \textit{I}   &   20.83 (0.09)   &   Keck/LRIS\\
2457983.5000   &   0.50   &   \textit{z}   &   17.67 (0.03)   &   Magellan/LDSS-s\\
2457984.4768   &   1.48   &   \textit{z}   &   17.62 (0.06)   &   Magellan/LDSS\\
2457984.4784   &   1.48   &   \textit{z}   &   17.61 (0.06)   &   Magellan/LDSS\\
2457984.4790   &   1.48   &   \textit{z}   &   17.61 (0.06)   &   Magellan/LDSS\\
2457986.5000   &   3.50   &   \textit{z}   &   18.38 (0.05)   &   Magellan/LDSS-s\\
2457988.4714   &   5.47   &   \textit{z}   &   19.08 (0.12)   &   Magellan/LDSS\\
2457990.4698   &   7.47   &   \textit{z}   &   19.87 (0.07)   &   Magellan/LDSS\\
2457991.4720   &   8.47   &   \textit{z}   &   20.40 (0.07)   &   Magellan/LDSS\\
2457995.5185   &   12.52   &   \textit{z}   &   $>$ 22.48   &   Magellan/IMACS\\
2457984.5022   &   1.50   &   \textit{J1}   &   17.32 (0.01)   &   Magellan/FourStar\\
2457985.4801   &   2.48   &   \textit{J1}   &   17.52 (0.01)   &   Magellan/FourStar\\
2457987.4705   &   4.47   &   \textit{Y}   &   18.08 (0.02)   &   du Pont/RetroCam\\
2457988.5000   &   5.50   &   \textit{Y}   &   18.33 (0.10)   &   du Pont/RetroCam\\
2457995.4600   &   12.46   &   \textit{J1}   &   $>$ 21.31    &   Magellan/FourStar\\
2457983.5000   &   0.50   &   \textit{J}   &   17.83 (0.15)   &   Magellan/FourStar\\
2457984.4813   &   1.48   &   \textit{J}   &   17.47 (0.01)   &   Magellan/FourStar\\
2457985.4625   &   2.46   &   \textit{J}   &   17.55 (0.01)   &   Magellan/FourStar\\
2457986.4728   &   3.47   &   \textit{J}   &   17.85 (0.01)   &   Magellan/FourStar\\
2457987.4986   &   4.50   &   \textit{J}   &   18.10 (0.02)   &   du Pont/RetroCam\\
2457990.4903   &   7.49   &   \textit{J}   &   18.99 (0.04)   &   du Pont/RetroCam\\
2457991.5000   &   8.50   &   \textit{J}   &   19.30 (0.10)   &   du Pont/RetroCam\\
2457995.4600   &   12.46   &   \textit{J}   &   $>$ 20.65   &   Magellan/FourStar\\
2457996.5030   &   13.50   &   \textit{J}   &   $>$ 21.16   &   Magellan/FourStar\\
2457983.4902   &   0.49   &   \textit{H}   &   18.26 (0.15)   &   Magellan/FourStar\\
2457984.4902   &   1.49   &   \textit{H}   &   17.52 (0.01)   &   Magellan/FourStar\\
2457985.4714   &   2.47   &   \textit{H}   &   17.57 (0.01)   &   Magellan/FourStar\\
2457987.4825   &   4.48   &   \textit{H}   &   17.83 (0.02)   &   du Pont/RetroCam\\
2457988.5000   &   5.50   &   \textit{H}   &   17.84 (0.20)   &   du Pont/RetroCam\\
2457992.4696   &   9.47   &   \textit{H}   &   19.21 (0.05)   &   du Pont/RetroCam\\
2457993.4838   &   10.48   &   \textit{H}   &   19.34 (0.09)   &   du Pont/RetroCam\\
2457994.4970   &   11.50   &   \textit{H}   &   19.64 (0.14)   &   NTT/SOFI\\
2457995.4600   &   12.46   &   \textit{H}   &   $>$ 20.30   &   Magellan/FourStar\\
2457996.4620   &   13.46   &   \textit{H}   &   $>$ 20.50   &   Magellan/FourStar\\
2457998.4697   &   15.47   &   \textit{H}   &   $>$ 20.50   &   Magellan/FourStar\\
2457983.5300   &   0.53   &   \textit{K$_s$}   &   18.41 (0.15)   &   Magellan/FourStar\\
2457984.4810   &   1.48   &   \textit{K$_s$}   &   17.61 (0.04)   &   Magellan/FourStar\\
2457985.4620   &   2.46   &   \textit{K$_s$}   &   17.55 (0.06)   &   Magellan/FourStar\\
2457995.4620   &   12.46   &   \textit{K$_s$}   &   19.36 (0.09)   &   Magellan/FourStar\\
2457995.4850   &   12.48   &   \textit{K$_s$}   &   19.32 (0.09)   &   NTT/SOFI\\
2457996.4780   &   13.48   &   \textit{K$_s$}   &   19.52 (0.09)   &   Magellan/FourStar\\
2457996.4900   &   13.49   &   \textit{K$_s$}   &   19.43 (0.09)   &   NTT/SOFI\\
2457998.4697   &   15.47   &   \textit{K$_s$}   &   20.23 (0.10)   &   Magellan/FourStar\\
2458001.4600   &   18.46   &   \textit{K$_s$}   &   20.81 (0.13)   &   Magellan/FourStar\\
2458001.4900   &   18.49   &   \textit{K$_s$}   &   20.93 (0.17)   &   Magellan/FourStar\\
\hline
\label{tab:Photometry}
\end{longtable}

\clearpage

\begin{table}
\caption{{\bf Photometric Properties of SSS17a.} Columns include the observed broad-band filter, the peak observed apparent magnitude (m$_{\rm{obs,max}}$), the peak observed absolute magnitude (M$_{\rm{obs,max}}$), the observed rise time (t$_{\rm{rise}}$), the decline rate measured in the first 5 days post-observed-maximum, the time for the light curve to decline by half of its peak flux (t$_{1/2,\rm{decline}}$), and the number of magnitudes the light curve declines in the first five days post-maximum ($\Delta$m$_5$). Times are given in rest-frame days since LVC trigger.  Uncertainties ($1\sigma$) are given in parentheses next to each measurement. Upper and lower limits are designated with the symbols ``$<$'' and ``$>$'', respectively. All magnitude measurements are listed on the AB scale. Apparent magnitudes have been corrected for Milky Way extinction. Errors in absolute magnitude account for an uncertainty in the distance to NGC\,4993 of 3 Mpc based on the range of distances listed in the NASA/IPAC Extragalactic Database.}
\begin{tabular}{@{}lcccccc}\hline\hline
Filter & m$_{\rm{obs,max}}$ & M$_{\rm{obs,max}}$ & t$_{\rm{rise}}$ &
Decline Rate & t$_{1/2,\rm{decline}}$ & $\Delta$m$_5$ \\
 & (mag) & (mag) & (day) & (mag day$^{-1}$) & (day) & (mag) \\ \hline
$K_s$ & 17.51 (0.05) & -15.46 (0.17) & 2.5 (0.5) &  & 3.11 (0.15) & 0.85 (0.06) \\
$H$ & 17.46 (0.02) & -15.51 (0.16) & 1.5 (0.5) & 0.10 (0.01) & 3.66 (0.95) & 0.88 (0.02) \\
$J$ & 17.37 (0.01) & -15.60 (0.16) & 1.5 (0.5) & 0.23 (0.01) & 3.41 (0.06) & 1.23 (0.03) \\
$Y$ & 17.20 (0.01) & -15.78 (0.16) & $<$1.5 & 0.25 (0.01) & 2.98 (0.09) & 1.26 (0.18) \\
$z$ & 17.45 (0.06) & -15.52 (0.17) & 1.0 (0.5) & 0.38 (0.01) & 2.00 (0.50) & 1.86 (0.11) \\
$i$ & 17.26 (0.15) & -15.71 (0.23) & $<$0.5 & 0.57 (0.01) & 1.78 (0.31) & 2.80 (0.19) \\
$r$ & 17.00 (0.14) & -15.97 (0.21) & $<$0.5 & 0.95 (0.02) & 1.12 (0.15) & 4.06 (0.30) \\
$V$ & 16.93 (0.15) & -16.04 (0.23) & $<$0.5 & 1.13 (0.04) & 0.79 (0.12) & 5.58 (0.37) \\
$g$ & 17.01 (0.15) & -15.96 (0.16) & $<$0.5 & 1.14 (0.02) & 0.62 (0.02) & 5.39 (0.13) \\
$B$ & 18.61 (0.06) & -14.35 (0.17) & $<$1.5 & 1.37 (0.06) & 0.51 (0.05) & 5.77 (0.35) \\
$u$ & 17.71 (0.07) & -15.26 (0.17) & $<$0.67 & 2.58 (0.31) & 0.30 (0.05) & \\
$w1$ & 18.74 (0.09) & -14.23 (0.18) & $<$0.67 & 2.65 (0.74) & 0.28 (0.07) &  \\
$m2$ & 20.28 (0.22) & -12.69 (0.27) & $<$0.67 & $>$ 3.2 &  &  \\
$w2$ & 20.32 (0.22) & -12.65 (0.27) & $<$0.67 & $>$ 2.3 &  &  \\
\hline
\end{tabular}
\end{table}


\begin{thebibliography}{10}

\bibitem{Abbott16:bbh}
B.~P. {Abbott}, {\it et~al.\/}, {\it Physical Review X\/} {\bf 6}, 041015
  (2016).

\bibitem{Abbott16:gw15}
B.~P. {Abbott}, {\it et~al.\/}, {\it Physical Review Letters\/} {\bf 116},
  061102 (2016).

\bibitem{Hughes2003}
S.~A. {Hughes}, D.~E. {Holz}, {\it Classical and Quantum Gravity\/} {\bf 20},
  S65 (2003).

\bibitem{Holz2005}
D.~E. {Holz}, S.~A. {Hughes}, {\it \apj\/} {\bf 629}, 15 (2005).

\bibitem{Nissanke2013}
S.~{Nissanke}, {\it et~al.\/}, {\it ArXiv: astro-ph/1307.2638\/}  (2013).

\bibitem{Phinney2009}
E.~S. {Phinney}, {\it Astro2010: The Astronomy and Astrophysics Decadal Survey,
  Science White Papers\/}, ArXiv: astro-ph/0903.0098 (2009).

\bibitem{Mandel2010}
I.~{Mandel}, R.~{O'Shaughnessy}, {\it Classical and Quantum Gravity\/} {\bf
  27}, 114007 (2010).

\bibitem{Paczynski1986}
B.~{Paczynski}, {\it \apjl\/} {\bf 308}, L43 (1986).

\bibitem{Eichler1989}
D.~{Eichler}, M.~{Livio}, T.~{Piran}, D.~N. {Schramm}, {\it \nat\/} {\bf 340},
  126 (1989).

\bibitem{Kelley2013}
L.~Z. {Kelley}, I.~{Mandel}, E.~{Ramirez-Ruiz}, {\it \prd\/} {\bf 87}, 123004
  (2013).

\bibitem{Fong2015}
W.~{Fong}, E.~{Berger}, R.~{Margutti}, B.~A. {Zauderer}, {\it \apj\/} {\bf
  815}, 102 (2015).

\bibitem{Li1998}
L.-X. {Li}, B.~{Paczy{\'n}ski}, {\it \apjl\/} {\bf 507}, L59 (1998).

\bibitem{Kulkarni2005}
S.~R. {Kulkarni}, {\it ArXiv: astro-ph/0510256\/}  (2005).

\bibitem{Metzger2010}
B.~D. {Metzger}, {\it et~al.\/}, {\it \mnras\/} {\bf 406}, 2650 (2010).

\bibitem{Roberts2011}
L.~F. {Roberts}, D.~{Kasen}, W.~H. {Lee}, E.~{Ramirez-Ruiz}, {\it \apjl\/} {\bf
  736}, L21 (2011).

\bibitem{Piran2013}
T.~{Piran}, E.~{Nakar}, S.~{Rosswog}, {\it \mnras\/} {\bf 430}, 2121 (2013).

\bibitem{Metzger2017}
B.~D. {Metzger}, {\it Living Reviews in Relativity\/} {\bf 20}, 3 (2017).

\bibitem{Shen2015}
S.~{Shen}, {\it et~al.\/}, {\it \apj\/} {\bf 807}, 115 (2015).

\bibitem{Burbidge1957}
E.~M. {Burbidge}, G.~R. {Burbidge}, W.~A. {Fowler}, F.~{Hoyle}, {\it Reviews of
  Modern Physics\/} {\bf 29}, 547 (1957).

\bibitem{Cameron1957}
A.~G.~W. {Cameron}, {\it \pasp\/} {\bf 69}, 201 (1957).

\bibitem{Qian2007}
Y.-Z. {Qian}, G.~J. {Wasserburg}, {\it \physrep\/} {\bf 442}, 237 (2007).

\bibitem{Arnould2007}
M.~{Arnould}, S.~{Goriely}, K.~{Takahashi}, {\it \physrep\/} {\bf 450}, 97
  (2007).

\bibitem{Tanvir2013}
N.~R. {Tanvir}, {\it et~al.\/}, {\it \nat\/} {\bf 500}, 547 (2013).

\bibitem{Berger2013}
E.~{Berger}, W.~{Fong}, R.~{Chornock}, {\it \apjl\/} {\bf 774}, L23 (2013).

\bibitem{GCN21509}
{LIGO/Virgo collaboration}, {\it GRB Coordinates Network\/} {\bf 21509} (2017).

\bibitem{GCN21513}
{LIGO/Virgo collaboration}, {\it GRB Coordinates Network\/} {\bf 21513} (2017).

\bibitem{Abbott17:ns}
B.~P. {Abbott}, {\it et~al.\/}, {\it Accepted to Physical Review Letters, DOI =
  10.1103/PhysRevLett.119.161101\/}  (2017).

\bibitem{GCN21529}
{One-Meter Two-Hemisphere (1M2H) collaboration}, {\it GRB Coordinates
  Network\/} {\bf 21529} (2017).

\bibitem{Coulter2017}
{Coulter et~al.}, {\it Science, this issue 10.1126/science.aap9811\/}  (2017).

\bibitem{GCN21551}
{Simon et~al.}, {\it GRB Coordinates Network\/} {\bf 21551} (2017).

\bibitem{GCN21547}
{Drout et~al.}, {\it GRB Coordinates Network\/} {\bf 21547} (2017).

\bibitem{Shappee2017}
B.~J. {Shappee}, {\it et~al.\/}, {\it Science, this issue
  10.1126/science.aaq0186\/}  (2017).

\bibitem{MM}
Materials and methods are available as supplementary materials.

\bibitem{Siebert17}
{Siebert et~al.}, {\it accepted to ApJL, 10.3847/2041-8213/aa905e\/}  (2017).

\bibitem{Arnett1982}
W.~D. {Arnett}, {\it \apj\/} {\bf 253}, 785 (1982).

\bibitem{Kasen2013}
D.~{Kasen}, N.~R. {Badnell}, J.~{Barnes}, {\it \apj\/} {\bf 774}, 25 (2013).

\bibitem{Perego2014}
A.~{Perego}, {\it et~al.\/}, {\it \mnras\/} {\bf 443}, 3134 (2014).

\bibitem{Fernandez2016}
R.~{Fern{\'a}ndez}, B.~D. {Metzger}, {\it Annual Review of Nuclear and Particle
  Science\/} {\bf 66}, 23 (2016).

\bibitem{Hotokezaka2013}
K.~{Hotokezaka}, {\it et~al.\/}, {\it \prd\/} {\bf 87}, 024001 (2013).

\bibitem{Fernandez2015}
R.~{Fern{\'a}ndez}, D.~{Kasen}, B.~D. {Metzger}, E.~{Quataert}, {\it \mnras\/}
  {\bf 446}, 750 (2015).

\bibitem{Wanajo2014}
S.~{Wanajo}, {\it et~al.\/}, {\it \apjl\/} {\bf 789}, L39 (2014).

\bibitem{Bovard17}
L.~{Bovard}, {\it et~al.\/}, {\it ArXiv: astro-ph/1709/09630\/}  (2017).

\bibitem{Kasen2015}
D.~{Kasen}, R.~{Fern{\'a}ndez}, B.~D. {Metzger}, {\it \mnras\/} {\bf 450}, 1777
  (2015).

\bibitem{Kilpatrick17}
{Kilpatrick et~al.}, {\it Science, this issue, 10.1126/science.aaq0073\/}
  (2017).

\bibitem{Elmhamdi2003}
A.~{Elmhamdi}, {\it et~al.\/}, {\it \mnras\/} {\bf 338}, 939 (2003).

\bibitem{Metzger14}
B.~D. {Metzger}, A.~L. {Piro}, {\it \mnras\/} {\bf 439}, 3916 (2014).

\bibitem{Metzger2009}
B.~D. {Metzger}, A.~L. {Piro}, E.~{Quataert}, {\it \mnras\/} {\bf 396}, 304
  (2009).

\bibitem{Gottlieb17}
O.~{Gottlieb}, E.~{Nakar}, T.~{Piran}, {\it ArXiv: astro-ph/1705.10797\/}  (2017).

\bibitem{Kappeler89}
F.~{Kappeler}, H.~{Beer}, K.~{Wisshak}, {\it Reports on Progress in Physics\/}
  {\bf 52}, 945 (1989).

\bibitem{Qian2000}
Y.-Z. {Qian}, {\it \apjl\/} {\bf 534}, L67 (2000).

\bibitem{Abbott16:review}
B.~P. {Abbott}, {\it et~al.\/}, {\it Living Reviews in Relativity\/} {\bf 19}
  (2016).

\bibitem{Yaron2012}
O.~{Yaron}, A.~{Gal-Yam}, {\it \pasp\/} {\bf 124}, 668 (2012).

\bibitem{Guillochon2017}
J.~{Guillochon}, J.~{Parrent}, L.~Z. {Kelley}, R.~{Margutti}, {\it \apj\/} {\bf
  835}, 64 (2017).

\bibitem{Rest2005}
A.~{Rest}, {\it et~al.\/}, {\it \apj\/} {\bf 634}, 1103 (2005).

\bibitem{Rest2014}
A.~{Rest}, {\it et~al.\/}, {\it \apj\/} {\bf 795}, 44 (2014).

\bibitem{Scolnic15}
D.~{Scolnic}, {\it et~al.\/}, {\it \apj\/} {\bf 815}, 117 (2015).

\bibitem{Dressler2006}
A.~{Dressler}, T.~{Hare}, B.~C. {Bigelow}, D.~J. {Osip}, {\it Society of
  Photo-Optical Instrumentation Engineers (SPIE) Conference Series\/} (2006),
  vol. 6269 of {\it Proc.SPIE}, p. 62690F.

\bibitem{IRAF}
IRAF is distributed by the National Optical Astronomy Observatory, which is
  operated by the Association of Universities for Research in Astronomy (AURA)
  under a cooperative agreement with the National Science Foundation.

\bibitem{Chambers2016}
K.~C. {Chambers}, {\it et~al.\/}, {\it ArXiv: astro-ph/1612.05560\/}  (2016).

\bibitem{Stetson1987}
P.~B. {Stetson}, {\it \pasp\/} {\bf 99}, 191 (1987).

\bibitem{Flewelling2016}
H.~A. {Flewelling}, {\it et~al.\/}, {\it ArXiv: astro-ph/1612.05243\/}  (2016).

\bibitem{Persson2013}
S.~E. {Persson}, {\it et~al.\/}, {\it \pasp\/} {\bf 125}, 654 (2013).

\bibitem{Skrutskie2006}
M.~F. {Skrutskie}, {\it et~al.\/}, {\it \aj\/} {\bf 131}, 1163 (2006).

\bibitem{Blanton2005}
M.~R. {Blanton}, {\it et~al.\/}, {\it \aj\/} {\bf 129}, 2562 (2005).

\bibitem{Krisciunas17}
K.~{Krisciunas}, {\it et~al.\/}, {\it ArXiv: astro-ph/1709.05146\/}  (2017).

\bibitem{GCN21550}
{Swift team}, {\it GRB Coordinates Network\/} {\bf 21550} (2017).

\bibitem{GCN21572}
{Swift team}, {\it GRB Coordinates Network\/} {\bf 21572} (2017).

\bibitem{Poole2008}
T.~S. {Poole}, {\it et~al.\/}, {\it \mnras\/} {\bf 383}, 627 (2008).

\bibitem{Breeveld2010}
A.~A. {Breeveld}, {\it et~al.\/}, {\it \mnras\/} {\bf 406}, 1687 (2010).

\bibitem{Smartt2015}
S.~J. {Smartt}, {\it et~al.\/}, {\it \aap\/} {\bf 579}, A40 (2015).

\bibitem{Buzzoni1984}
B.~{Buzzoni}, {\it et~al.\/}, {\it The Messenger\/} {\bf 38}, 9 (1984).

\bibitem{Moorwood1998}
A.~{Moorwood}, J.-G. {Cuby}, C.~{Lidman}, {\it The Messenger\/} {\bf 91}, 9
  (1998).

\bibitem{GCN21582}
{Lyman et~al.}, {\it GRB Coordinates Network\/} {\bf 21582} (2017).

\bibitem{Oke1995}
J.~B. {Oke}, {\it et~al.\/}, {\it \pasp\/} {\bf 107}, 375 (1995).

\bibitem{Steidel2004}
C.~C. {Steidel}, {\it et~al.\/}, {\it \apj\/} {\bf 604}, 534 (2004).

\bibitem{Landolt1992}
A.~U. {Landolt}, {\it \aj\/} {\bf 104}, 340 (1992).

\bibitem{LPipe}
Http://www.astro.caltech.edu/~dperley/programs/lpipe.html.

\bibitem{Henden2016}
A.~A. {Henden}, {\it et~al.\/}, {\it VizieR Online Data Catalog: II/336\/}
  (2016).

\bibitem{Freedman2001}
W.~L. {Freedman}, {\it et~al.\/}, {\it \apj\/} {\bf 553}, 47 (2001).

\bibitem{Schlafly2011}
E.~F. {Schlafly}, D.~P. {Finkbeiner}, {\it \apj\/} {\bf 737}, 103 (2011).

\bibitem{Cardelli1989}
J.~A. {Cardelli}, G.~C. {Clayton}, J.~S. {Mathis}, {\it \apj\/} {\bf 345}, 245
  (1989).

\bibitem{Drout2014}
M.~R. {Drout}, {\it et~al.\/}, {\it \apj\/} {\bf 794}, 23 (2014).

\bibitem{Brown2016}
P.~J. {Brown}, A.~{Breeveld}, P.~W.~A. {Roming}, M.~{Siegel}, {\it \aj\/} {\bf
  152}, 102 (2016).

\bibitem{Metzger2008}
B.~D. {Metzger}, A.~L. {Piro}, E.~{Quataert}, {\it \mnras\/} {\bf 390}, 781
  (2008).
  
\bibitem{Barnes2016}
J.~{Barnes}, D.~{Kasen}, M.-R. {Wu}, G.~{Mart{\'{\i}}nez-Pinedo}, {\it \apj\/}
  {\bf 829}, 110 (2016).

\bibitem{Bauswein2013}
A.~{Bauswein}, S.~{Goriely}, H.-T. {Janka}, {\it \apj\/} {\bf 773}, 78 (2013).

\bibitem{Rosswog2014}
S.~{Rosswog}, O.~{Korobkin}, A.~{Arcones}, F.-K. {Thielemann}, T.~{Piran}, {\it
  \mnras\/} {\bf 439}, 744 (2014).

\bibitem{Wheeler15}
J.~C. {Wheeler}, V.~{Johnson}, A.~{Clocchiatti}, {\it \mnras\/} {\bf 450}, 1295
  (2015).

\bibitem{Clocchiatti97}
A.~{Clocchiatti}, J.~C. {Wheeler}, {\it \apj\/} {\bf 491}, 375 (1997).

\end{thebibliography}
\end{document}